\documentclass[useAMS,usenatbib]{mn2e}

\usepackage{times}
\usepackage{deluxetable}
\usepackage{amsmath}
\usepackage{graphicx}  
\usepackage{url}

\newcommand{\myemail}{roskar@physik.uzh.ch}

\def\aj{AJ}
\def\araa{ARA\&A}
\def\apj{ApJ}
\def\apjl{ApJL}
\def\apjs{ApJS}

\def\aap{A\&A}

\def\mnras{MNRAS}

\def\ksk{km/s/kpc}
\def\dj{$\Delta j_z$~}

\def\jjmax{$j_z/j_c(E)$~}

\title[Radial Migration in Disk Galaxies I]{Radial Migration in Disk Galaxies I: Transient Spiral Structure and Dynamics}

\author[R. Ro\v{s}kar et al.]{Rok Ro\v{s}kar$^{1,2}$\thanks{\myemail}, 
Victor P. Debattista$^{3,4}$,
Thomas R. Quinn$^2$,
James Wadsley$^5$
\\
$^1$Institute for Theoretical Physics, University of Z\"{u}rich, Winterthurerstrasse 190, CH-8057 Z\"{u}rich, Switzerland\\
$^2$Astronomy Department, University of Washington, Box 351580, Seattle, WA 98195, USA\\
$^3$Jeremiah Horrocks Institute, University of Central Lancashire, Preston, PR1 2HE, UK\\
$^4$RCUK Fellow\\
$^5$Department of Physics and Astronomy, McMaster University, Hamilton, ON, L8S 4M1, Canada\\
}

\begin{document}

\maketitle

\begin{abstract}

We seek to understand the origin of radial migration in spiral
galaxies by analyzing in detail the structure and evolution of an
idealized, isolated galactic disk. To understand the redistribution of
stars, we characterize the time-evolution of properties of spirals
that spontaneously form in the disk. Our models unambiguously show
that in such disks, single spirals are unlikely, but that a number of
transient patterns may coexist in the disk. However, we also show that
while spirals are transient in amplitude, at any given time the disk
favors patterns of certain pattern speeds. Using several runs with
different numerical parameters we show that the properties of spirals
that occur spontaneously in the disk do not sensitively depend on
resolution. The existence of multiple transient patterns has large
implications for the orbits of stars in the disk, and we therefore
examine the resonant scattering mechanisms that profoundly alter
angular momenta of individual stars. We confirm that the corotation
scattering mechanism described by \citet{Sellwood:2002} is responsible
for the largest angular momentum changes in our simulations.

\end{abstract}


%
%

\section{Introduction}

As stellar galactic disks form and evolve, the processes governing
their development leave behind a multitude of traces. Massive stars
pollute the interstellar medium through supernova explosions and
stellar winds, endowing subsequent generations of stars with distinct
chemical signatures. Similarly, dynamical perturbations, whether they
are secular (spiral arms or bars) or due to the hierchical build-up of
mass, sculpt the kinematic properties of stellar
populations. Together, these signatures provide a ``fossil record'' of
a disk's evolution through cosmic time.

Galaxy formation models attempt to match their predictions to this
fossil record and attempt to reconstruct the disk's history. For the
past thirty years, an implicit assumption has been made in such
modeling: stars remain in the same part of the disk forever
\citep[e.g.][]{Tinsley:1975, Matteucci:1989, Boissier:1999,
  Chiappini:2001}. This simple assumption carries enormous power as it
allows one to reconstruct the time evolution of a given quantity (such
as metallicity and star formation rate) at a particular radius from
present-day, single epoch observations. Such ``static'' modeling has
been successful in illuminating several critical aspects of disk
formation such as the need for infall of low metallicity gas to solve
the G-dwarf problem \citep{Tinsley:1975}. Over the past several years,
however, this assumption of stars remaining near their birth radii has
been firmly shaken by the realization that rapid stellar migrations of
several kpc are possible \citep[][SB02 hereafter]{Sellwood:2002}.

The idea that stars may not remain near their birth radii is not new.
\citet{Wielen:1977} suggested that stellar orbits diffuse in velocity
space consequently inducing a drift in the galactocentric radius for
an ensemble of stars. This was followed up by \citet{Wielen:1985} and
applied to the determination of the Sun's birthplace
\citep{Wielen:1996}. In \citet{Wielen:1977} the diffusion is a
relatively slow process driven by random scattering by giant molecular
clouds - for observationally-constrained velocity dispersions,
variations in galactocentric radius of at most few kpc are
expected. Instead, in the mechanism described by SB02, migrations of
\emph{several} kpc may take place in a few hundred Myr due to very
efficient exchange of angular momentum at the corotation resonance
(CR).

This exchange of energy and angular momentum at the CR occurs without
changing the orbital radial actions therefore retaining the orbital
circularity.  In addition, the mechanism operates most efficiently for
stars on the most circular orbits. These two properties lead to the
peculiar result that migrations of several kpc in a MW-size disk are
possible without substantial accompanying increase in random
motion. The increase in velocity dispersion with age of stars (see
e.g. \citealt{Holmberg:2009} for the solar neighborhood) constrains
heating processes and thereby also the magnitude of spiral
perturbations, so the ability of the CR to redistribute stars
substantially without heating excessively is a critical aspect of the
CR mixing process. Because the SB02 mechanism involves the CR, the
spiral amplitudes must be transient if \emph{redistribution} rather
than \emph{trapping} is to occur. In the absence of other
perturbations, a steady spiral traps stars on horseshoe orbits -
however, if a spiral's amplitude grows and decays on a timescale
comparable to one half the libration period of a horseshoe orbit, the
spiral will merely deposit the star on the other side of the resonance
and vanish before pulling it back (SB02).

Unlike steady spirals that can heat the disk only at the Lindblad
resonances \citep{Lynden-Bell:1972}, transient spirals can also heat
the disk away from the principal resonances \citep{Barbanis:1967,
Binney:1988, Jenkins:1990}. Consequently, a single spiral can heat the
disk enough to prevent any further asymmetric structure formation in
just a few rotations \citep{Sellwood:1984}. Star formation provides a
natural cooling mechanism, continually peppering the disk with young
stars born with the galactic velocities of their parent gas
clouds. \citet{Sellwood:1984} found that in the presence of such
cooling, transient spirals keep the disk in a quasi-stable state at a
Toomre $Q\sim2$, which allows for their continual regeneration.

The realization that stars in galactic disks may migrate radially
across significant distances has in recent years completely changed
the discourse on spiral galaxy evolution. Following SB02, several
other theoretical works have further illuminated the complexities that
radial mixing introduces for stellar population
studies. \citet{Lepine:2003} investigated the effect of corotation
scattering on the disk metallicity gradient.  \citet[][hereafter
R08a]{Roskar:2008} showed that radial migration could drastically
alter the stellar population properties of outer disks and provided an
explanation for the observed gradients across the surface brightness
profile break in NGC4244 \citep{de-Jong:2007}.  R08a predicted that
disks with broken exponential profiles should show an inflection in
the mean age profile corresponding to the break radius. This has
subsequently been confirmed indirectly by surface photometry
\citep{Bakos:2008, Azzollini:2008} and directly by integral-field
spectroscopy (\citealt{Yoachim:2010}, Yoachim et al. 2012, in press)
and resolved-star counts (Radburn-Smith 2012, submitted), providing
further evidence that radial mixing occurs in external galaxies. 

Using the same simulations, \citet[][hereafter R08b]{Roskar:2008a}
extended the analysis of R08a and investigated the repercussions of
stellar migration for a range of stellar population studies: the solar
neighborhood age-metallicity relation (AMR) and metallicity
distribution function (MDF), the evolution of metallicity gradients,
and the reconstruction of star formation histories from present-day
observations of stellar populations in external galaxies. They found
that $>50\%$ of stars on mostly circular orbits in the solar
neighborhood of their model have come from elsewhere and that
migration flattens the AMR and broadens the MDF. They also found that
the reconstruction of a star formation history becomes problematic in
the presence of migration especially at large radii where the mass in
migrated stars approaches or exceeds the cumulative amount of stars
formed in-situ. R08ab therefore established the notion that regardless
of the method used to observe a galactic disk (unresolved photometry,
photometry of resolved stars, spectroscopy of individual stars in the
MW) radially migrated stars substantially alter the combined
properties of the observed sample, thereby possibly strongly biasing
the outcome of any subsequent modelling.

\citet{Schonrich:2009, Schonrich:2009a} included a probabilistic
prescription of radial migration in a chemical evolution model
constrained by Milky Way observables and found that, similarly to
R08b, the MDF is broadened and the AMR flattened as a result of
migration. They found also that most obervational properties and
peculiarities of the thick disk can be explained by radial migration.
\citet{Loebman:2011} explored the formation of the thick disk via
radial migration in the $N$-body models of R08ab and similarly found
that simulated trends tend to agree qualitatively with observed thick
disk properties from the Sloan Digital Sky Survey (SDSS)
\citep{Ivezic:2008,
  Lee:2011}. \citet{Minchev:2010,Minchev:2011,Minchev:2012} argued
that apart from corotation scattering as presented by SB02, combined
effects of multiple nearby resonances from several different patterns
may also be important in driving the redistribution of stars. Their
results implied that substantial mixing is possible even if the spiral
structure is not transient. \citet{Brunetti:2011} measured the
diffusion coefficients in idealized bar-unstable disks and found the
CR of the bar to be driving the diffusion, though their models were
collisionless and as such did not allow for recurrent spiral activity.

Understanding radial migration is particularly relevant at the present
time because of the upcoming surveys designed to study the detailed
structure of the Milky Way disk. Studies using data from the SDSS
(e.g. \citealt{Juric:2008, Ivezic:2008}) and its follow-up surveys
such as SEGUE \citep{Yanny:2009, DeJong:2010, Lee:2011}, as well as
results using data from RAVE \citep{Wilson:2011, Ruchti:2011} have
already pushed the current models to their limits in trying to explain
the various observed trends and interdependencies in the thick and
thin disks. In the coming years, spectroscopic surveys, such as HERMES
\citep{Freeman:2008} and APOGEE \citep{Prieto:2008}, will obtain
high-resolution spectroscopy of millions of stars, allowing finally
for ``chemical tagging'' of stars into their birth clusters
\citep{Freeman:2002}. Such observations will in principle allow for a
direct measurement of stellar radial migration in the Milky Way
\citep{Bland-Hawthorn:2010}. At the same time, the \textit{Gaia}
mission will provide a complete 6D map of a 10 kpc sphere centered on
the Sun \citep{Perryman:2001}, and follow-up spectroscopic surveys
will yield vast amounts of complementary chemical abundance
data. Finally, the Large Synoptic Survey Telescope
\citep{LSST-Science-Collaborations:2009} will similarly provide
ground-based photometry of the entire sky. To distill these vast data
sets and apply them to galactic archeology, a clear understanding of
the radial mixing processes is essential.

In this paper, we explore in detail the properties of self-propagating
spiral structure and the causes of the resulting radial migration in
simulations from R08ab and \citet{Loebman:2011}. We first quantify the
spiral structure that forms spontaneously in our simulations and use
this analysis as a basis for understanding the causes of radial
migration. We also perform a set of numerical tests to explore the
robustness of our results to choices of numerical parameters and
stochasticity. The Paper is organized as follows: in
$\S$~\ref{sec:sim} we discuss the details of our models; in
$\S$~\ref{sec:spiral_anatomy} we quantify the spiral structure in our
fiducial simulation; in $\S$~\ref{sec:angular_momentum_corotation} we
explore the causes of radial mixing; in $\S$~\ref{sec:numerical_tests}
we present several tests of the effects of numerical parameters on the
generation of spiral structure; finally, we state our conclusions in
$\S$~\ref{sec:conclusions}.

%
%

\section{Simulations}
\label{sec:sim}

The initial conditions for all of the runs presented here are
generated as described in \citet{Kaufmann:2006, Kaufmann:2007}, and
consist of spherical distributions of dark matter (DM) and gas
following an NFW \citep{Navarro:1997} density profile. The random
velocities of the DM particles are initialized to ensure equilibrium
by means of the distribution function obtained from an inversion of
the NFW density profile \citep{Kazantzidis:2004b}. The gas is
initialized to the same mass distribution with a temperature profile
to yield an approximate hydrostatic equilibrium. The gas is also given
a spin consistent with values obtained from collisionless cosmological
simulations \citep{Bullock:2001, Maccio:2007}, i.e.  $\lambda = (J/G)
\sqrt{|E|/M^5} = 0.039$, where $J$ is the total angular momentum, $E$
is the total energy of the system and $M$ is the mass. The angular
momentum follows a radial profile of $j\propto r$, where $j$ is the
specific angular momentum. In the fiducial run we use $10^6$ particles
per component, resulting in DM particle mass of $10^6
\mathrm{~M_{\odot}}$ and initial gas mass of $1.4\times10^5
\mathrm{~M_{\odot}}$.

The simulations were run using the code \textsc{GASOLINE}
\citep{Wadsley:2004}, a hydrodynamics extension of the parallel
multi-stepping $N$-body code \textsc{PKDGRAV} \citep{Stadel:2001}. The
tuning of the sub-grid star formation and feedback prescriptions are
described extensively in \citet{Stinson:2006}. Here we summarize the
important features. A gas particle becomes eligible for star formation
when its density exceeds $0.1 \text{ cm}^{-1}$ and its temperature
dips below 15,000~K. If eligible, the gas particle converts some of
its mass into a star particle at a rate given by
\begin{equation}
\frac{\text{d}\rho_{\star}}{\text{d}t}=c_{\star}\frac{\rho_{gas}}{t_{dyn}},
\end{equation}
where $c_{\star}$ is the star formation efficiency (set to 0.05 in all
of the simulations), $\rho_{gas}$ is the gas density and $t_{dyn}$ is
the local dynamical time. Star particles form with 1/3 of the initial
gas particle mass (in the fiducial simulation this translates to star
particle masses of $4.7\times10^4\mathrm{~M_{\odot}}$) and to avoid
unreasonably small particle masses, gas particle masses are limited to
1/5 of their initial mass. When a particle crosses this threshold its
mass is distributed among neighboring particles, resulting in an
overall decrease of the number of gas particles with time. Each star
particle represents an entire stellar population and therefore a
spectrum of stellar masses described by the Miller-Scalo initial mass
function \citep{Miller:1979}. The evolution of massive stars is
followed and a feedback cycle is initiated to reflect the explosion of
type II supernovae. At typical particle masses of $\sim10^4
M_{\odot}$, the supernova energy is injected into the ISM ``in bulk'',
i.e. individual explosions are not modeled. The effect of the
supernova explosions is modeled on the sub-grid level as a blastwave
propagating through the ISM. We track ISM metal enrichment from type
II and type Ia supernovae as well as from AGB stars. 

The complete set of parameters used for the simulations discussed in
this Paper is listed in Table~\ref{table:runs}. Our fiducial run, has
been studied extensively and we used it previously for results
presented in R08ab. The baryonic particles in the fiducial run use a
softening length $h_s = 50$~pc. We tested the softening dependence of
our results with runs S1, S3 and S4 ($h_s = 25, 100$~and 500~pc
respectively). We also tested the effect of varying particle numbers
in runs R1, R3 and R4 with $0.5\times10^6$, $2\times10^6$ and
$4\times10^6$ particles in each component respectively. We ran a
further test of the effects of the mass resolution of the DM halo by
running a simulation with 10 times the number of DM particles (run
SDM). All runs were carried out for 10 Gyr.

The opening angle $\Theta$ used for simplifying gravity calculations
was 0.7 for all runs. The code uses a multistepping scheme with the
condition that a particle's timestep $\Delta t_{grav} = \eta
(\epsilon/a)^{1/2}$, where $\eta=0.175$, $\epsilon$ is the
gravitational softening length, and $a$ is the acceleration. For the
gas particles, the time step must also satisfy $\Delta t_{gas} =
\eta_{courant} h/[(1+\alpha)c + \beta\mu_{max}]$, where
$\eta_{courant} = 0.4$, $h$ is the SPH smoothing length, $\alpha=1$ is
the shear coefficient and $\beta=2$ is the viscosity
coefficient. $\mu_{max}$ is described in \citet{Wadsley:2004}. The SPH
quantities were calculated on a kernel using 32 nearest neighbors. The
cooling in all runs is calculated without taking into account the
metal content of the gas. Additional simulations where we vary the
physical parameters of the initial halos (mass, angular momentum) will
be presented in Paper~II of this series.

\begin{deluxetable}{lcccccc}
\centering
\tablecaption{Simulation Parameters\tablenotemark{a}}
\tabletypesize{\small} 
\tablewidth{0pt} 
\tablehead{
\colhead{Name} & 
\colhead{$N_{gas}$} & 
\colhead{$N_{dark}$} & 
\colhead{$h_s$ [kpc]} 
}
\startdata 
fiducial\tablenotemark{b} & $10^6$ & $10^6$ & 0.05 \\
S1 & $10^6 $ & $10^6$ & 0.025 \\ 
S3 & $10^6 $ & $10^6$ & 0.1 \\ 
S4 & $10^6 $ & $10^6$ & 0.5 \\ 
SDM & $10^6 $ & $10^7$ & 0.05 \\
T2\tablenotemark{c} & $10^6 $ & $10^6$ & 0.05\\ 
T3\tablenotemark{c} & $10^6 $ & $10^6$ & 0.05\\
R1 & $5\times10^5 $ & $5\times10^5$ & 0.05 \\ 
R3 & $2\times10^6 $ & $2\times10^6$ & 0.05 \\ 
R4 & $4\times10^6 $ & $4\times10^6$ & 0.05 \\ 
R1-T2\tablenotemark{d} & $5\times10^5 $ & $5\times10^5$ & 0.05 \\ 
\enddata

\tablenotetext{a}{All runs have $M_{vir} =
  10^{12}\mathrm{~M_{\odot}}$, $\lambda = 0.039$ and $c = 8.0.$}

\tablenotetext{b}{This is our fiducial run that was presented
  previously in R08ab.}

\tablenotetext{c}{Runs T2 and T3 are identical to the fiducial run
  S2/R2 but use different random seeds when generating the initial
  conditions.}

\tablenotetext{d}{Run R1-T2 is identical to run R1 but initialized
  with a different random seed.}

\label{table:runs}
\end{deluxetable}

The strength of these idealized models lies in the fact that there is
no initial stellar component, so none of the properties of the disk
are chosen \textit{a priori}. As soon as the simulation begins, the
gas is allowed to cool in the potential well of the DM, and as it
reaches densities and temperatures conducive to star formation (as set
by our star formation parameters) it spawns stars. Due to the spin,
the gas naturally settles into a rotationally-supported disk in the
center of the halo, giving rise to a galactic disk comprised of a mix
of stars and gas. Thus, although all of the simulations presented here
are idealized in the sense that the disks evolve in isolation the
simulations also depart sharply from most other work using idealized
disks because we do not \textit{construct} equilibrium disk models, as
is usually done in studies focusing on dynamical effects
\citep[e.g.][]{Debattista:2006}.  Instead, our models grow disks
spontaneously without any biases regarding disk structure or stellar
population properties. The spontaneous growth is particularly
important for the development of self-limiting structure and
subsequent secular evolution \citep[e.g.][]{Sellwood:1984}. We are
therefore able to follow the temporal evolution of stellar populations
and make direct comparisons between simulated properties and observed
galaxies with unprecedented detail.

Real disks of course do not form in such a simple way, but our
simulations are meant only to mimic the formation of disks after the
last major merger. For massive galaxies ($M \geq 10^{11} M_{\odot}$)
disk formation is not dominated by filamentary accretion after
z$\sim$2 \citep{Keres:2005, Brooks:2009}. Minor mergers may influence
the morphology of the disk \citep[e.g.][]{Kazantzidis:2008}, but the
majority of the disk fuel comes from quiescent cooling of gas from the
halo. 

By limiting the processes included in our models, we can perform a
cleaner analysis\footnote{We analyze the simulations using our own IDL
  routines as well as the publicly available Python-based
  \texttt{pynbody} package:\\
  \url{http://code.google.com/p/pynbody/}} of the disk
dynamics. Because of lower computational cost compared to
fully-cosmological simulations, we are also able to attain higher
resolution. At the fiducial resolution, for example, the disks have
$\sim$ 2.5 million star particles by 10 Gyr - typical hydrodynamic
cosmological simulations of $10^{12} M_{\odot}$ galaxies have a few
times fewer disk particles\footnote{Cosmological simulations may have
  larger numbers of particles in the system, but most comprise the
  other galactic components like the bulge and halo and relatively few
  are found in the disk component.}. However, by allowing our disks to
form spontaneously without a pre-defined stellar component, we can
follow the evolution of stellar populations self-consistently
throughout the simulation.

%
%
%

\section{Anatomy of Spirals}
\label{sec:spiral_anatomy}
We begin by dissecting the spiral structure in our fiducial simulation
to show that spirals are responsible for the migration of stars. For
several decades, idealized $N$-body simulations have produced
self-limiting, transient spirals arising from various sources of
instability, such as Poisson noise, discontinuities in the
distribution function \citep{Sellwood:1991}, or stimulation by the
central bar \citep{Debattista:2006}. Studies of the evolution of
transient spiral structure have largely been confined to the realm of
carefully-tuned $N$-body experiments, sometimes with an addition of
gas hydrodynamics or an ad-hoc cooling mechanism
\citep{Sellwood:1984}. Cooling is essential for the persistence of
transient spirals because spirals are naturally self-destructive: when
they form, they also very efficiently heat the disk. Controlled and
carefully tuned $N$-body experiments are vital for understanding the
orbit response of the underlying disk to perturbing wave propagation,
but do not adequately address how spontaneously recurring patterns may
affect stars in a real disk. SB02 (their Figs. 10-12) provide a hint
and show that an unconstrained collisionless simulation produces an
array of spirals, identified by local maxima in the power spectrum of
m=2 perturbations.  In this section, we seek to expand on such
analysis by also considering the temporal evolution of patterns.

\begin{figure}
\centering
\includegraphics[width=3.5in]{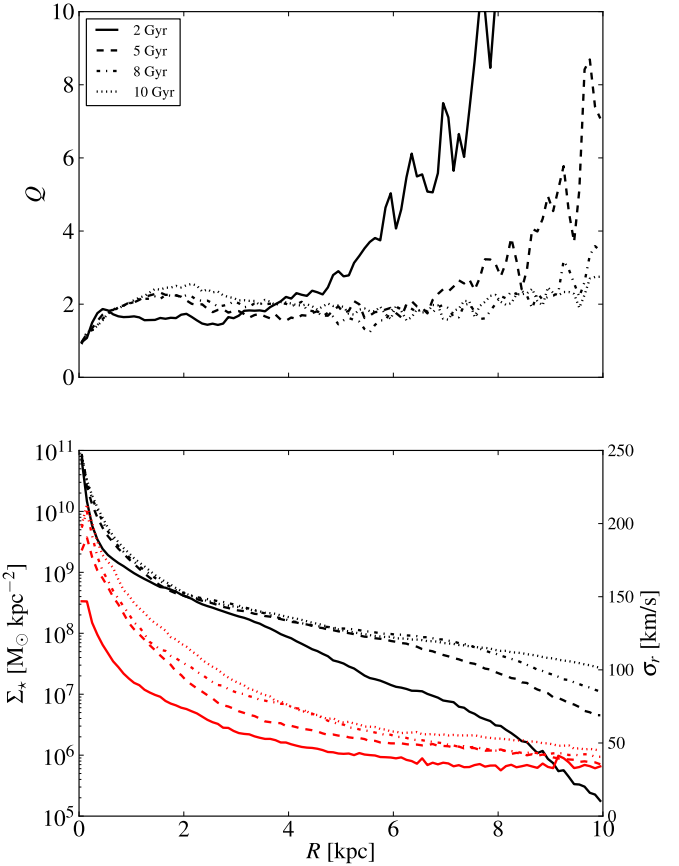}
\caption{{\bf Top:} Toomre $Q$ parameter as a function of radius at
  four different times in the fiducial run. {\bf Bottom:} Stellar
  surface density (black) and radial velocity dispersion (red). Line
  types correspond to the same times as in the upper panel.}
\label{fig:toomre_evol}
\end{figure}

In our simulations, the stellar disks are cooled naturally by on-going
star formation. Stars are initially on kinematically cool orbits
inherited from their parent gas particles, continuously infusing the
disk with a population susceptible to supporting instabilities. The
spiral structure manifests, as we show below, in a complicated array
of patterns and pattern speeds that evolve as the disk
grows. Figure~\ref{fig:toomre_evol} shows the Toomre $Q\equiv
\sigma_{R} \kappa/3.36 \Sigma G$ parameter as a function of radius at
several different times, where $\sigma_R$ is the radial velocity
dispersion of the stars, $\kappa$ is the epicyclic frequency, $\Sigma$
is the surface density of stars, and $G$ is the gravitational
constant. The bottom panel shows $\Sigma$ and $\sigma_R$
separately. The recurring spirals are possible because the disk
establishes a marginally stable state, with $Q\sim2$ in the main part
of the disk at all times.

In order to study the spiral structure, we center the system on the
potential minium and divide the particles into concentric equal-width
radial bins. We then expand the stellar particle distribution in a
Fourier series given by
\begin{equation}
\label{eq:fourier_expansion}
\Sigma(r,\phi) = \sum^{\infty}_{m=0} c_m(r)e^{[-i m \phi_m(r)]},
\end{equation}
where $r$ is the radius, $m$ is the pattern multiplicity, and
$\phi_m(r)$ is the phase of the $m$-th mode at radius $r$. The complex
coefficients $c_m(r)$ are given by
\begin{equation}
\label{eq:fourier_coefficients}
c_m(r) = \frac{1}{M(r)}\sum^N_{j=1} m_j e^{i m \phi_j},
\end{equation}
where $M(r)$ is the total mass in the radial bin, $m_j$ is the
particle mass, $\phi_j$ is the angle between the particle's
position vector and the $x$-axis, and $N$ is the total number of particles
in the bin satisfying $r < r_j < r+\delta r$. Strictly speaking, this type of
decomposition will identify any type of azimuthal m-fold symmetry, but
our assumption is that strong components identified in this way with low
$m$ will be due either to a bar or spirals. The bar can be separated
from spirals by virtue of its constant phase as a function of radius;
however, beyond the central few kpc we can reasonably expect this
method to identify spirals.

%

\begin{figure*}
\centering
\includegraphics[width=6in]{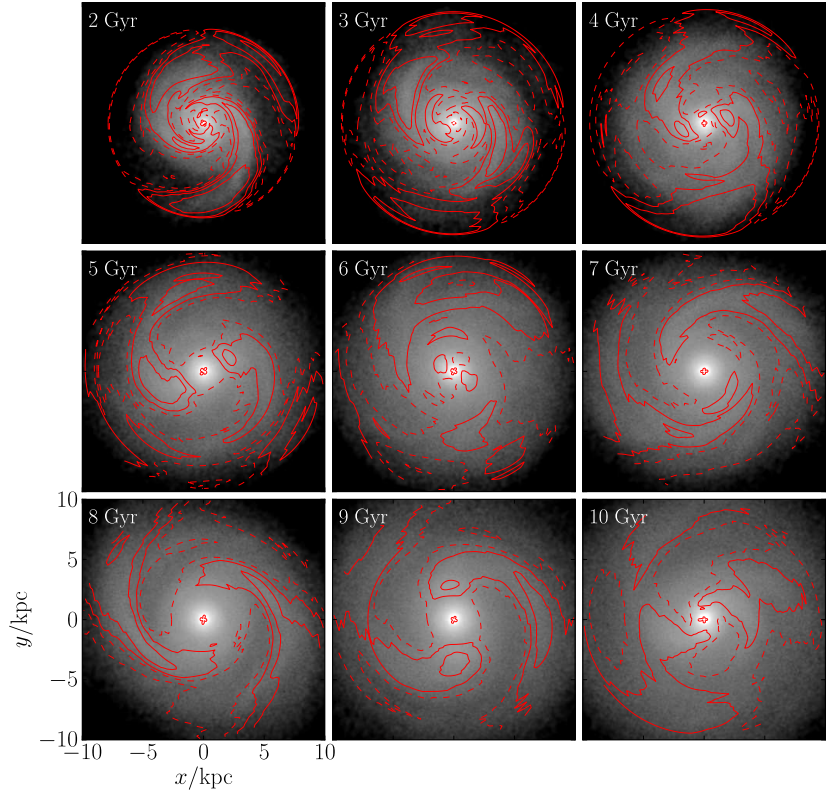}
\caption{Stellar density maps for the fiducial run at several
  times. Contours of overdensities reconstructed from the Fourier
  coefficients for m=1 to m=4 are overlaid in red. Contours are drawn
  at the -50, -20, -5, 5, 20, and 50 per cent (under-)over-densities;
  negative contours are shown with dashed lines. We also require that
  each radial bin has at least 1000 particles to avoid noise that may
  arise in regions with low particle numbers. All panels are the same
  physical scale.}
\label{fig:stellar_density_fourier_maps}
\end{figure*}

%

Figure~\ref{fig:stellar_density_fourier_maps} shows stellar density
maps with contours of m=1 through m=4 Fourier components, obtained using
equations \ref{eq:fourier_expansion} and
\ref{eq:fourier_coefficients}, demonstrating that our method can
reliably identify and account for the underlying disk structure. It is
also apparent from Figure~\ref{fig:stellar_density_fourier_maps} that
there are often distinct spirals present at different radii and that
each spiral often has a narrowly-defined peak overdensity. The
overdensities we measure vary with time but the normalized amplitude
of the m=2 mode $A_2 \equiv |c_2| \leq 0.4$ at all times for all runs,
compatible with the observations of \citet{Rix:1995}. Overdensities at
the level of a few per cent are difficult to see by eye in the stellar
surface density map, but are easily identified with a Fourier
decomposition.

\begin{figure}
\centering
\includegraphics[width=3.2in]{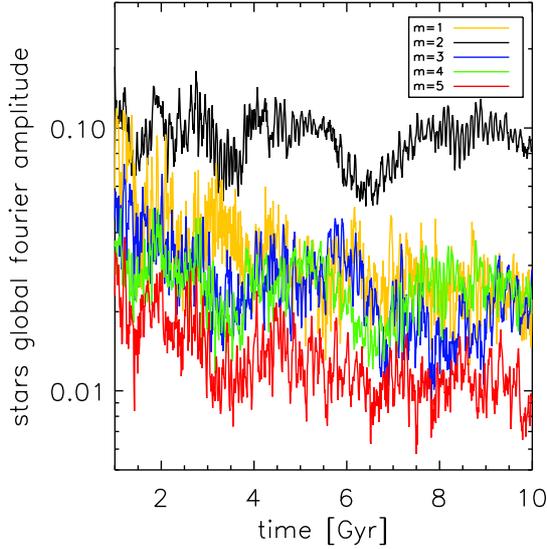}
\caption{Global Fourier amplitudes of the stellar particles for m=1
  through m=5 Fourier components.}
\label{fig:global_fourier}
\end{figure}

In a spontaneously-evolving disk, we expect patterns of all
multiplicities to exist, and they all could perturb the orbits of
stars. In Figure~\ref{fig:global_fourier} we show the global Fourier
amplitude as a function of time for m=1 through m=5 Fourier components. The m=2
mode dominates at all times in this simulation. We therefore 
focus our dynamical analysis on the m=2 mode, though we cannot 
exclude that the other components also contribute some small 
fraction to the angular momentum redistribution in the disk. 

%

The greatest impact of spiral structure on the orbits of stars occurs
at resonances, which are narrow regions in phase space satisfying
$m(\Omega_{\phi} - \Omega_p) = l\kappa$, where $m$ is the pattern
multiplicity, $\Omega_{\phi}$ and $\kappa$ are the azimuthal and
radial frequencies of the orbit and $\Omega_p$ is the spiral's pattern
speed. If $m=2$, setting $l$ to $\pm 1$ gives the inner and outer
Lindblad resonances (ILR and OLR), and $l=0$ corresponds to the
corotation resonance (CR). To identify resonances, we must therefore
obtain $\Omega_p$, which can be achieved using our Fourier
decomposition.

We can estimate the instantaneous $\Omega_p$ at radius $r$ simply by
performing numerical differentiation $\Omega_p =
\frac{\partial}{\partial t} \phi_m(r)$ or, equivalently, obtaining a
linear fit to $\phi_m(r)$ as a function of time. This method is
reliable if there is a single steady perturbation at the radius in
question, and studies of bars usually employ this method to recover
the bar pattern speed (e.g. \citealt{Debattista:2000,
  Dubinski:2009a}). We find, however, that for our purposes this is
insufficient because spirals of varying strengths and pattern speeds
are present at all radii, and this method only ever identifies the
most prominent pattern.

To reliably identify multiple pattern speeds in the disk, we require a
different method. Ideally, the method would not only allow us to
identify patterns at a given time interval, but it would also give us
information about the evolution of these patterns with time. However,
because the coefficients $c_m(r)$ are calculated at each output, they
comprise a time series and we can obtain a further discrete Fourier
transform of this series, given by 
\[
C_{k,m}(r) = \sum^{S-1}_{j=0} c_j(r,m) w_j e^{2\pi ijk/S} \qquad k = 0,...,S-1,
\]
\citep{Press:1992}, which yields the Fourier coefficients at discrete
frequencies $\Omega_k$, and $c_j(r,m)$ is the coefficient at time $t =
t_0 + j\Delta t$, radius $r$ and multiplicity $m$. $S$ is the number
of samples, in this case the number of outputs at which we evaluate
the structure. We use a Gaussian window function 
\[
w_j(x) = e^{-(x-S/2)^2/(S/4)^2},
\] 
to avoid high-frequency spectral leakage (using other window functions
such as the Hanning window has no appreciable effect). The frequency
sampling is determined by the length of the baseline and $\Delta t$,
the time between samples (outputs),
\[
\Omega_k = 2\pi\frac{k}{S\Delta t}m \qquad k=0,1...,\frac{S}{2},
\]
where $\Omega_{Ny} \equiv \Omega_{k=S/2}$ is the Nyquist frequency. We
are here faced with a choice between spectral resolution and the
ability to identify ``instantaneous'' pattern speeds. If the baseline
used is too long, the power spectrum will show a large number of
patterns not all of which are important for the disk at any particular
time. We find that at our timestep resolution of 10 Myr, using a 1 Gyr
(100 timesteps) baseline gives satisfactory spectra, though the
resolution is still rather coarse at $\sim 3$~\ksk. This sampling rate
gives a Nyquist frequency of $\Omega_{Ny} = 153$~\ksk~for $m=2$
structure.

\begin{figure}
\centering

\includegraphics[width=3.5in]{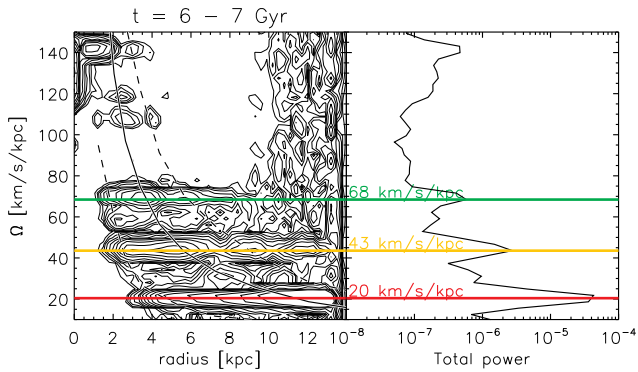}

\caption{{\bf Left:} Power spectrum of m=2 frequencies as a function
  of radius in the disk, for the time interval 6-7 Gyr.  The solid
  black line marks $\Omega_c$, while the dashed black lines mark
  $\Omega_c \pm \kappa/2$. The three main spiral patterns are
  identified by colored horizontal lines.  {\bf Right: } The global
  frequency spectrum. Contour levels vary in logarithm from
  $5\times10^{-10}$ to $1\times 10^{-6}$.}
\label{fig:frequencies}
\end{figure}

We can now construct a power spectrum at each radius given by
\[
P(\Omega_k,r)= \frac{1}{W} \left[|C_k(r)|^2 + |C_{S-k}(r)|^2\right] \quad k = 1,2,...\frac{S}{2}-1,
\]
where $W \equiv S\sum^S_{j=0}w_j$ is the normalization factor taking
into account the windowing function \citep{Press:1992}.  We assume
that the periodic changes will be due to the phase, rather than
amplitude variations, thus giving us pattern speeds
\citep{Sellwood:1986}. The resulting combined power spectrum across
the whole disk gives us information about the strengths of patterns as
a function of radius, and is shown in the
Figure~\ref{fig:frequencies}. The contours in the panels on the left
show the frequency power spectrum as a function of radius,
while the right panel shows the integrated power over the whole disk
at each frequency.

%

%

The pattern speeds in the given segment of the disk evolution can now
be obtained by extracting the peaks from the radially integrated
frequency spectrum. We measure the peaks by fitting Gaussians to the
local maxima in the integrated spectrum. To reliably identify
overlapping features in the spectrum we use an iterative procedure,
which progressively removes the most prominent peaks identifying
lesser peaks on subsequent passes. In this way, we identify the three
most important patterns. We use the Gaussian fit parameters to
estimate peak amplitudes and pattern speeds. The horizontal lines in
the left panel of Figure~\ref{fig:frequencies} correspond to the
pattern speeds obtained in this way and illustrate that this method
can recover the major pattern speeds well. The solid and dashed black
lines show $\Omega_c$, the circular frequency, and $\Omega_c \pm
\kappa/2$ - ILR, CR, and OLR resonances exist for stars on nearly
circular orbits where the pattern speed lines intersect one of these
loci, from left to right respectively.

Figure~\ref{fig:frequencies} reveals the richness of spiral structure
in a spontaneously evolving disk. The range of radii that may be
influenced by resonances from any of these patterns spans essentially
the entire disk. However, Figure~\ref{fig:frequencies} only tells us
about the patterns present in a given time interval - we would like to
know how these patterns evolve with time.

\begin{figure}
\centering

\includegraphics[width=3.5in]{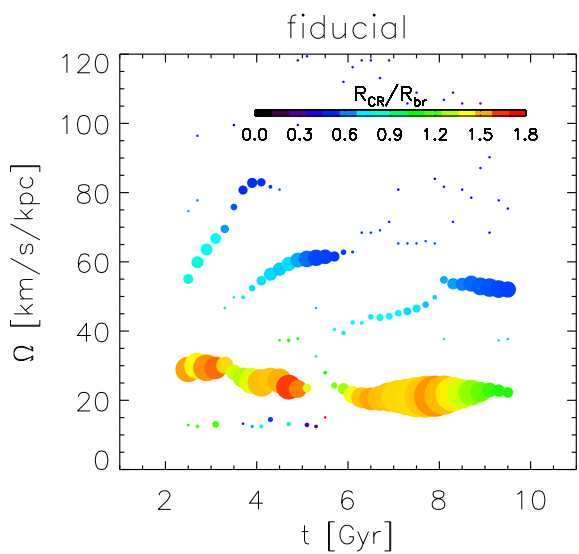}

\caption{Perturbation frequencies in the disk as a function of time
  obtained via the windowed FFT (see text). Point sizes reflect the
  normalized amplitude of each perturbation. Color denotes the ratio
  of corotation radius to the break radius.}
\label{fig:patspeed_sequence_fiducial}
\end{figure}

To extract time information from our data, we repeat the above
procedure at equally-spaced time intervals. We show such a result in
the Figure~\ref{fig:patspeed_sequence_fiducial}. The dominant
frequencies are shown as a function of time, identified using the
procedure outlined above, with individual point sizes reflecting the
normalized amplitude of each perturbation. Each point on the plot uses
a baseline of 1 Gyr of data to generate the power spectrum as
discussed above, so that at each time $t$ the patterns shown reflect
disk evolution from $t-1 \text{~Gyr}$ to $t$. Hence, adjacent points
represent overlapping data and the point sizes reflect a time-averaged
amplitude. The instantaneous amplitudes vary on much shorter
timescales as the spirals flicker in and out of existence.

%

%

It is striking that the perturbation at $\Omega_p \sim20-30$~\ksk~
dominates through most of the disk's lifetime. This perturbation is
very steady except for a brief decrease at 5.5 Gyr and a subsequent
strong growth peaking at 7.5 Gyr. The brief period of decreased power
can also be identified in the panels of
Figure~\ref{fig:stellar_density_fourier_maps}, which show that between
5-6 Gyr the patterns become significantly
disordered. Figure~\ref{fig:patspeed_sequence_fiducial} also shows
that perturbations with larger frequencies are significantly more
short-lived. We can identify three separate strong perturbations with
$\Omega_p > 50$~\ksk, the most persistent of which survives for $\sim
2$ Gyr.

The time evolution of pattern speeds and amplitudes shown in
Figure~\ref{fig:patspeed_sequence_fiducial} provides a new look at the
temporal evolution of the disk. We double-checked that the signal
extracted using the Windowed FFT (WFFT) method is indeed real by
performing an independent analysis using a continuous wavelet
transform (CWT), because it is specifically tailored to recover the
time-evolution of frequencies in a signal. The CWT yields
qualitatively identical results to the WFFT - however, we found it
less cumbersome to quantify the results (i.e. determine pattern speeds
and amplitudes) using the WFFT.

While Figure~\ref{fig:patspeed_sequence_fiducial} shows the evolution
of frequencies, it does not allow one to infer the amplitude
variations of individual patterns because the WFFT is performed on 0.5
Gyr intervals. Therefore, while we do see variation in the sizes of
the points, corresponding to amplitude changes, there is a possibility
that we are hiding shorter time-scale fluctuations. 

We attempt to reconstruct the amplitude variation of individual
patterns directly by using a band-pass filter in frequency space and
then using the inverse Fourier transform to obtain the time series of
Fourier coefficients representing the density perturbation. The
band-pass is centered roughly on the desired frequency and is 10
km/s/kpc wide. The choice of width is somewhat arbitrary, but we find
that the patterns are reasonably well separated with this
choice. 

\begin{figure}
\centering

\includegraphics[width=3.2in]{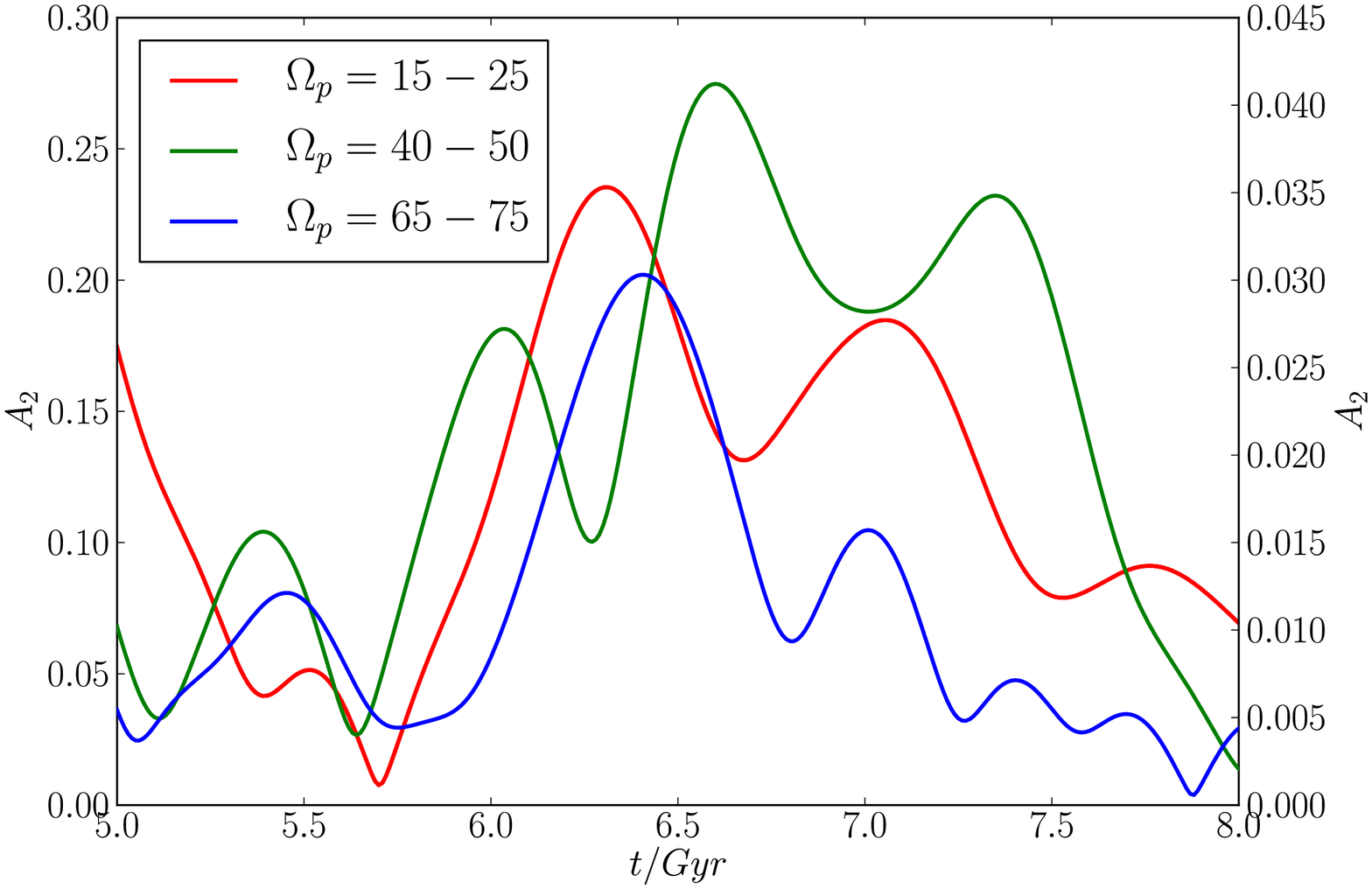}

\caption{Amplitude variation as a function of time for the three main
  patterns identified in Figure~\ref{fig:frequencies}. The amplitudes
  are obtained by using a band-pass of 10 km/s/kpc centered on each
  frequency and at a radius corresponding to the corotation radius at
  7 Gyr (i.e. identified using the power spectrum of
  Figure~\ref{fig:frequencies}. Because the amplitudes of the faster
  patterns are quite different, we use the left y-axis for the 20-30
  km/s/kpc pattern (red line), and the right y-axis for the other two
  (green and blue lines).}

\label{fig:spiral_amps}
\end{figure}

Figure~\ref{fig:spiral_amps} shows the resulting amplitude variations
for the three principal patterns found in
Figure~\ref{fig:frequencies}.  The oscillatory amplitude transience of
spirals on Gyr timescales in our run is clearly visible. Note that
apart from the slowest pattern, the pattern speeds are also evolving
on $\sim$~Gyr timescales (see
Figure~\ref{fig:patspeed_sequence_fiducial}). However, while the
evolving pattern speeds mean that resonances associated with these
patterns are sweeping through the disk, our system does not produce a
multitude of strong features of a wide range of pattern speeds at any
given time. Thus, the transience in the system described here might be
somewhat different from that found in other works.

%

%

\section{Angular Momentum Exchanges at Corotation in the Simulations}
\label{sec:angular_momentum_corotation}

In the previous section, we demonstrated that the disk in our
simulation harbors a variety of spiral patterns of various pattern
speeds. We now discuss how these patterns influence the orbits of
stars. In particular, we try to understand whether the radial
migration in our growing, unconstrained disk can be explained by the
CR scattering described in SB02 alone, or whether we must invoke more
complex mechanisms to explain the observed behavior.


\begin{figure*}
\centering
\includegraphics[width=6in]{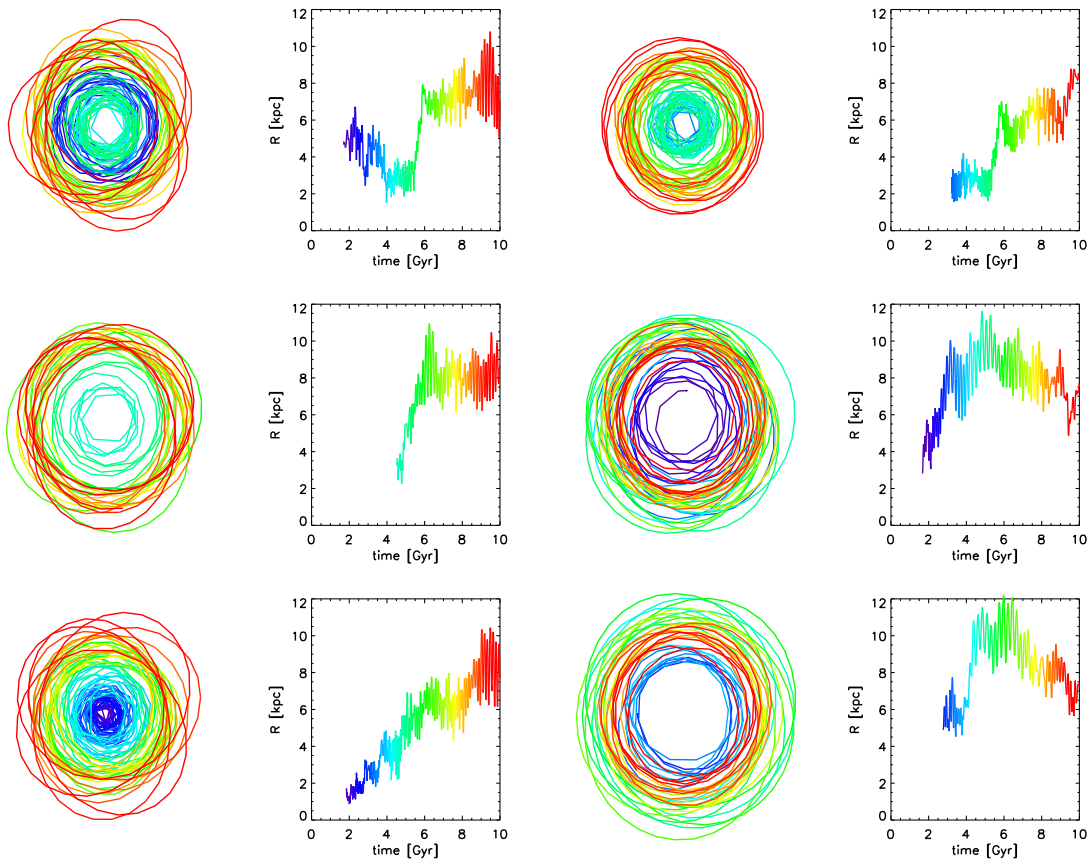}
\caption{Some examples of orbits from the fiducial simulation. Orbits
are colored according to time to make the temporal evolution more
apparent, with blue corresponding to early and red to late
times. Stars can migrate inwards and outwards very rapidly without
substantially increasing their eccentricities. }
\label{fig:redist_plots}
\end{figure*}

%

Figure~\ref{fig:redist_plots} illustrates the diversity of individual
stellar orbital histories in the disk. These orbits were chosen from a
random set of particles that are found beyond 3 kpc at the end of the
simulation, in order to illustrate the range of possible orbital
histories. It is clear from these examples that stars may migrate
rapidly while retaining a nearly circular orbit, and that a circular
orbit does not imply a radially static history. In fact the
instantaneous appearance of an orbit says very little about the star's
history, as it is evidently possible to even circularize a
significantly eccentric orbit (see, for example, the upper-right
panel).


\begin{figure}
\centering
\includegraphics[width=3.2in]{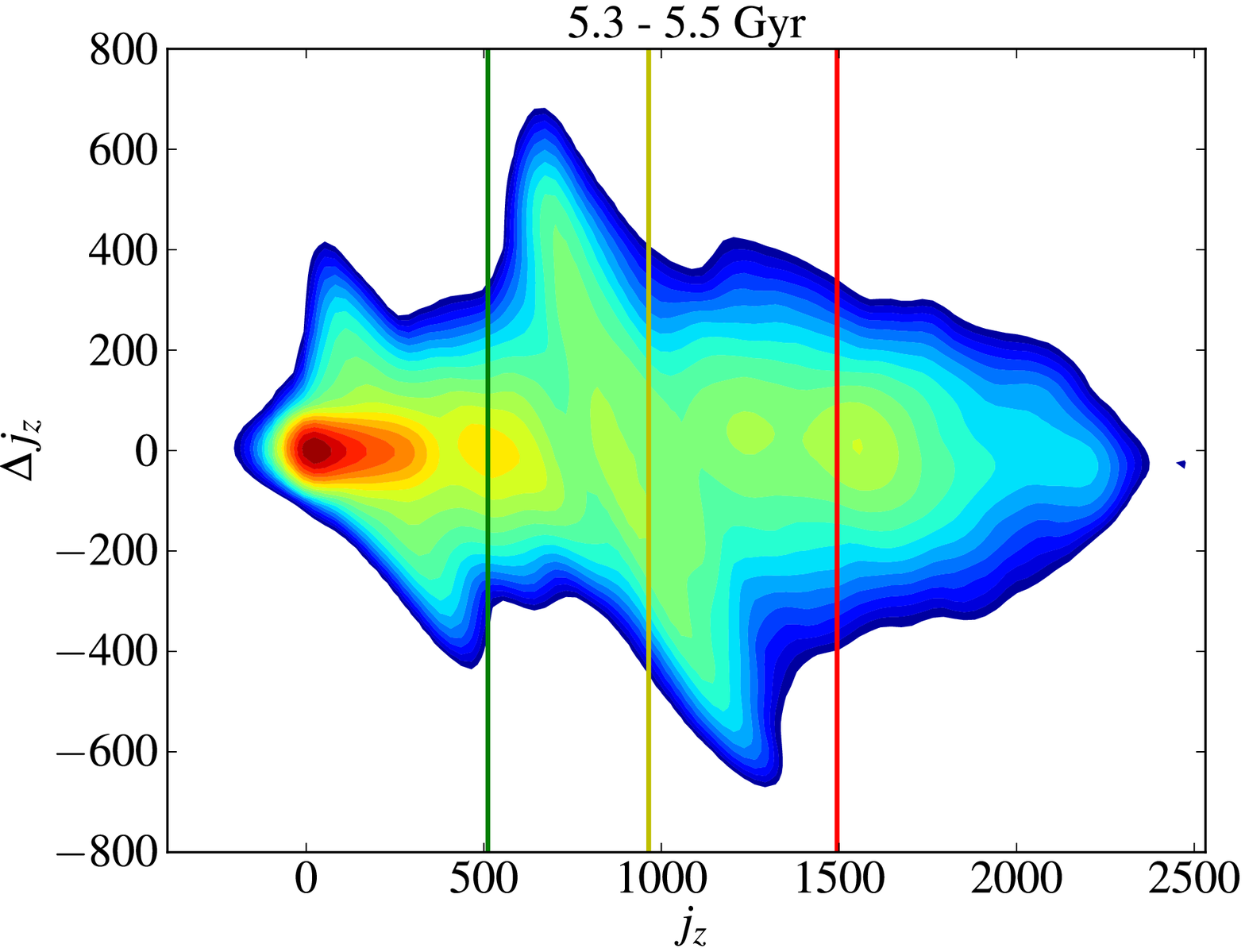}
\includegraphics[width=3.2in]{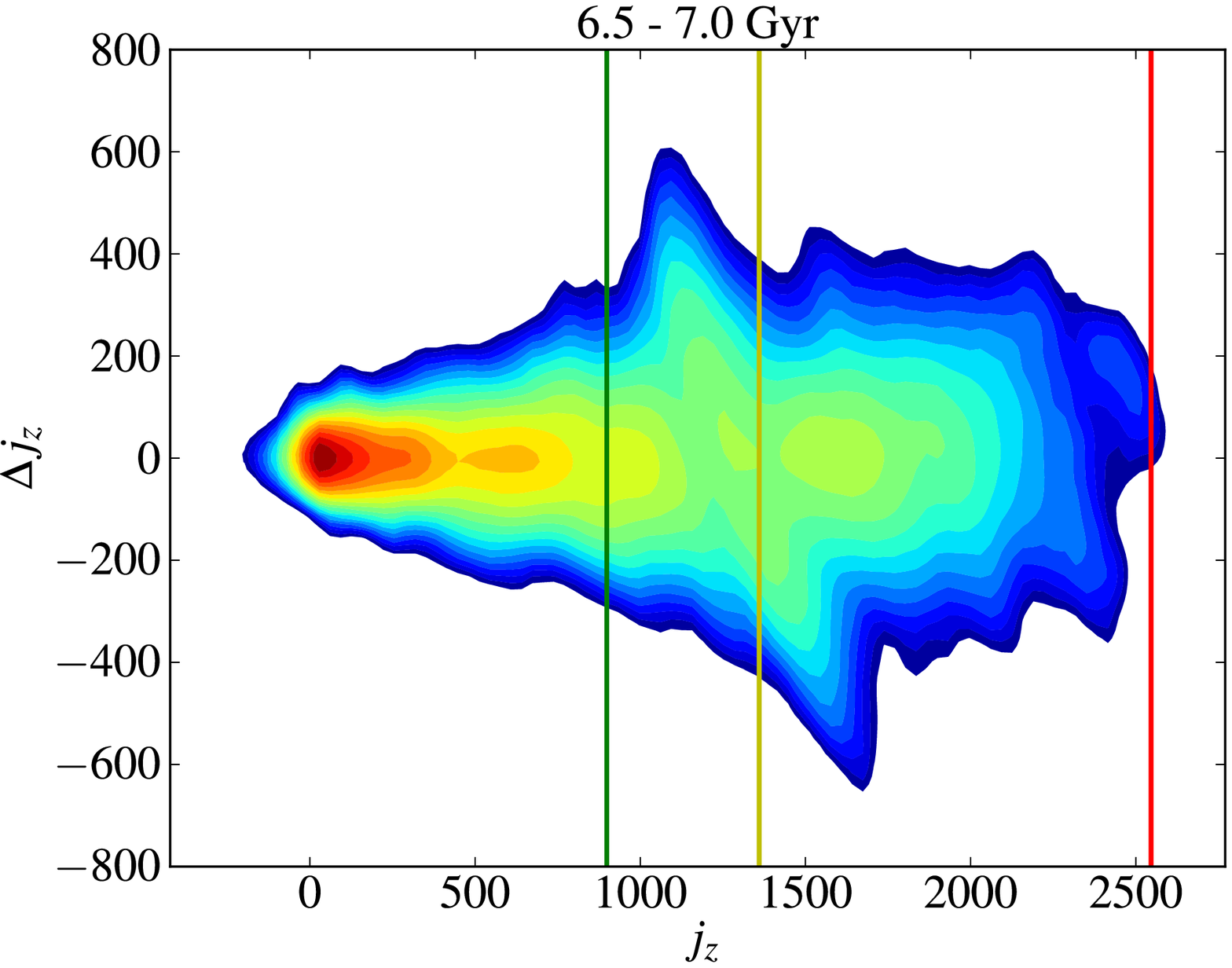}
\caption{Distributions of $\Delta j_z$ given initial $j_z$ for all
  particles in the simulation between 5.3-5.5 Gyr (top panel) and
  6.5-7.0 Gyr (bottom panel). The corotation resonances of the main
  perturbations found at each time are marked by vertical lines, with
  the colors in the top panel corresponding to the colors in
  Figure~\ref{fig:frequencies}. The logarithmically-spaced contour
  levels are the same for both panels and correspond to mass density.}
\label{fig:dj_j_tot}
\end{figure}

%

In Figure~\ref{fig:dj_j_tot} we show the changes in angular momentum
as a function of initial particle angular
momentum.\footnote{Throughout the paper we use the notation that
  specific angular momentum is lower-case $j$ and total angular
  momentum is capital $J$} For the bottom panel, the perturbations
dominating the disk during this time are the same ones we identified
in Figure~\ref{fig:frequencies}. Note that the baseline used to obtain
the frequency power spectrum is 1 Gyr, but the timescale on which we
consider \dj vs. $j_z$ is 0.5 Gyr, where $j_z =
\left[\mathbf{r}\times\mathbf{v}\right]_z$ and $\Delta j_z = j_z(t_2)
- j_z(t_1)$. If we use a 1 Gyr baseline (which corresponds to $> 4$
rotation periods at 5 Gyr and 8 kpc) we find that the signature of
spiral scattering in the \dj vs. $j_z$ plane becomes too smeared to be
useful. On the other hand, a 0.5 Gyr baseline is insufficient to
reliably identify pattern speeds in the power spectrum, so we are
forced to accept this small inconsistency. The higher frequency
perturbations may evolve considerably during $\sim 1 \text{~Gyr}$, but
the resonance loci are not a particularly steep function of $R$
(Figure~\ref{fig:frequencies}) so we can still obtain approximate
locations of resonances. In Figure~\ref{fig:spiral_amps} we can see
that during 6.5-7 Gyr we may expect the disk to be largely dominated
by the growth of the ~45 km/s/kpc pattern. However, during the 5.3-5.5
Gyr time interval, we see that the two faster patterns are growing at
the same time. It is also apparent from
Figure~\ref{fig:patspeed_sequence_fiducial} that around this time the
faster pattern dominates the overall perturbation amplitude in the
main part of the disk. We therefore choose this second time interval
to look for nonlinear effects stemming from resonant coupling of
multiple perturbations \citep[e.g.][]{Quillen:2003, Minchev:2010}.

In both cases, the dominant pattern causes the characteristic
anti-symmetric shape in $\Delta j_z$ vs $j_z$ associated with CR orbit
swapping (SB02). The few dominant patterns we identify cannot explain
the full structure seen in either panel, however. In
Figure~\ref{fig:frequencies} we can see that some amount of power is
present in subtler patterns not picked up by our automated algorithm,
such as at $\sim$~30~\ksk~and $\sim$~60~\ksk. Some of the structure in
Figure~\ref{fig:frequencies} must therefore also be due to those
patterns, which may be in the process of growing or fading at either
end of the time interval we consider. Another possibility is that
patterns of higher multiplicity contribute to some of the angular
momentum exchange. The prominent feature seen in the very inner part
of the disk in ther bottom panel is a short-lived weak bar
instability. Note the scale of $\Delta j_z$: in both panels the
changes caused by the dominant pattern approach 50\% of $j_z$. The
changes are especially drastic in the top panel, given the short time
interval we are considering.

\subsection{Corotation Crossing and Chaos}
\label{subsec:chaos}

\begin{figure*}
\centering
\includegraphics[width=6in]{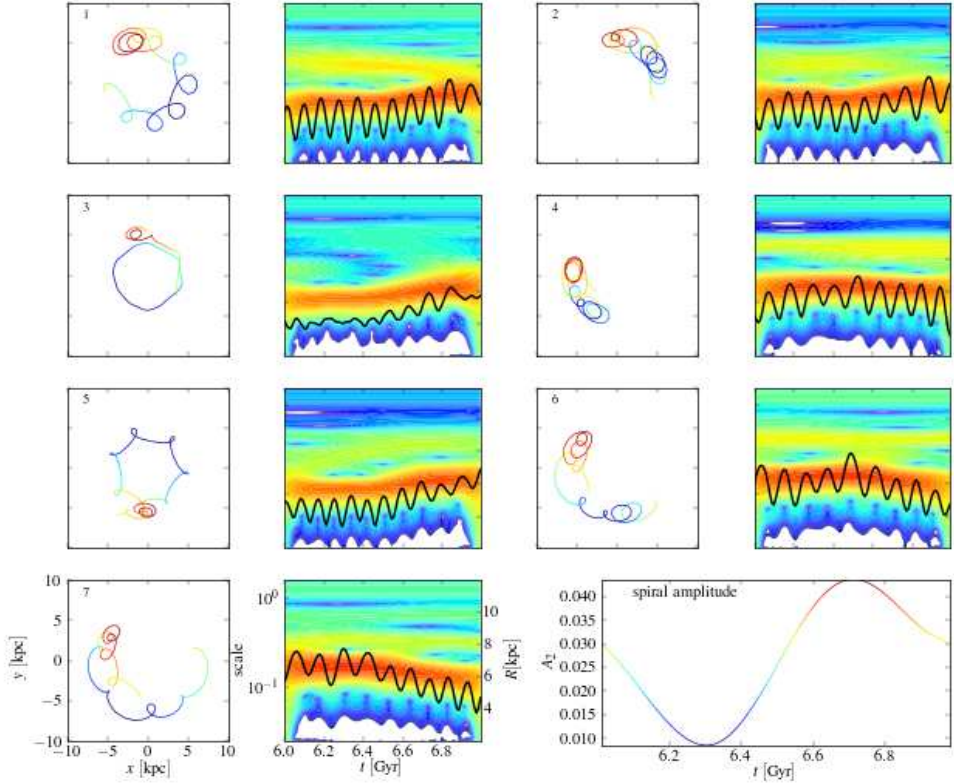}
\caption{{\bf Columns 1 and 3:} A selection of randomly selected
  orbits for a subsample of top 10 percent of migrators on either side
  of the CR for the 43 km/s/kpc pattern in the time interval 6.5-7
  Gyr. The orbits are in the frame corotating with the spiral. The
  color corresponds to the relative spiral amplitude, as shown in the
  bottom right panel. All boxes measure 20 kpc in $x$ and $y$. The
  direction of rotation is counter-clockwise. Note that the time of CR
  crossing is easily identified as the orbit reverses direction. All
  CR crossings happen around the time of the spiral amplitude peak and
  are spatially correlated. {\bf Columns 2 and 4:} The wavelet power
  spectrum of the $x$-component of the orbits is shown in color
  contours. Note that most of the features are horizontal with few
  bends and vary smoothly, indicating lack of significant chaotic
  evolution. The solid black lines show the radius $R$ in the
    $xy$ plane as a function of time. See the bottom left panels for
    axis scales.}
\label{fig:horseshoes}
\end{figure*}

A \emph{steady} spiral perturbation does not result in angular
momentum exchange at the CR because stars are trapped on horseshoe
orbits. If the spiral is transient, however, the trapping does not
occur and stars can traverse from one side of the CR to another. We
look for such horseshoe orbits during the time interval 6.5-7 Gyr
among the particles with $|\Delta j_z| > 200$ (roughly the top 10
percent of migrators) around the CR with the 43 km/s/kpc pattern shown
in the bottom panel of Figure~\ref{fig:dj_j_tot}. We plot the orbits
of a random subset of these particles in a frame corotating with the
spiral in columns 1 and 3 of Figure~\ref{fig:horseshoes}. The orbits
are colored according to the relative strength of the 43 km/s/kpc
spiral at any given time, shown in the bottom right panel. The radius
as a function of time is shown as a black line in columns 2 and 4. In
the corotating frame, a particle crossing the CR also reverses
direction. All particles cross the CR at almost the same time, near
the peak of the spiral amplitude, but the orbits are otherwise largely
unperturbed. Furthermore, all inward and outward migrations are
respectively correlated in azimuth. For example, see the outward
migrators in panels 1, 2 and 3 and the inward migrators in panels 6
and 7. This hints at the CR being responsible for the migration
because one expects the particles trailing the spiral arm to
preferentially be pulled outward and vice versa for those particles
leading the arm.

\begin{figure}
\centering
\includegraphics[width=3.2in]{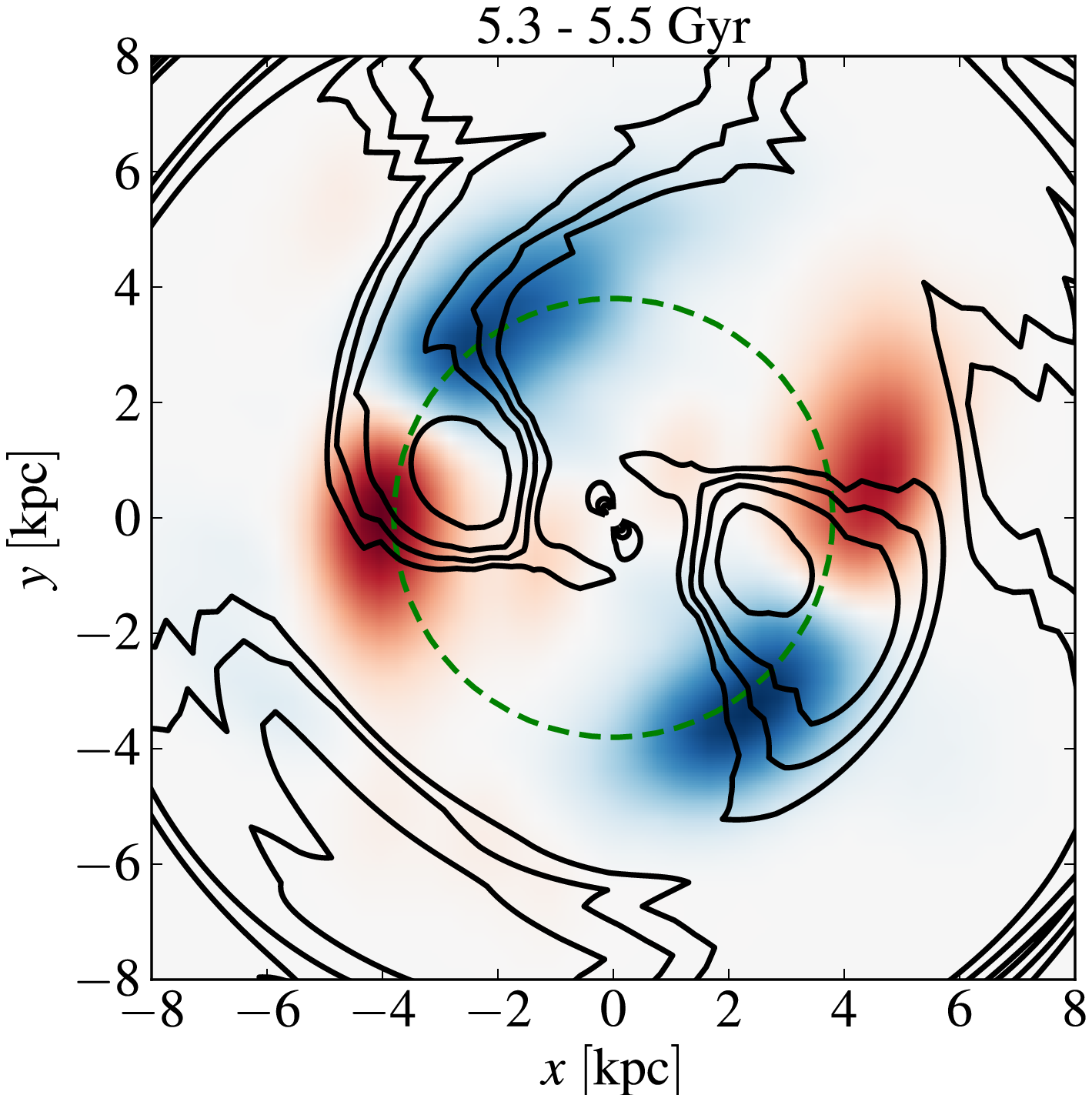}
\includegraphics[width=3.2in]{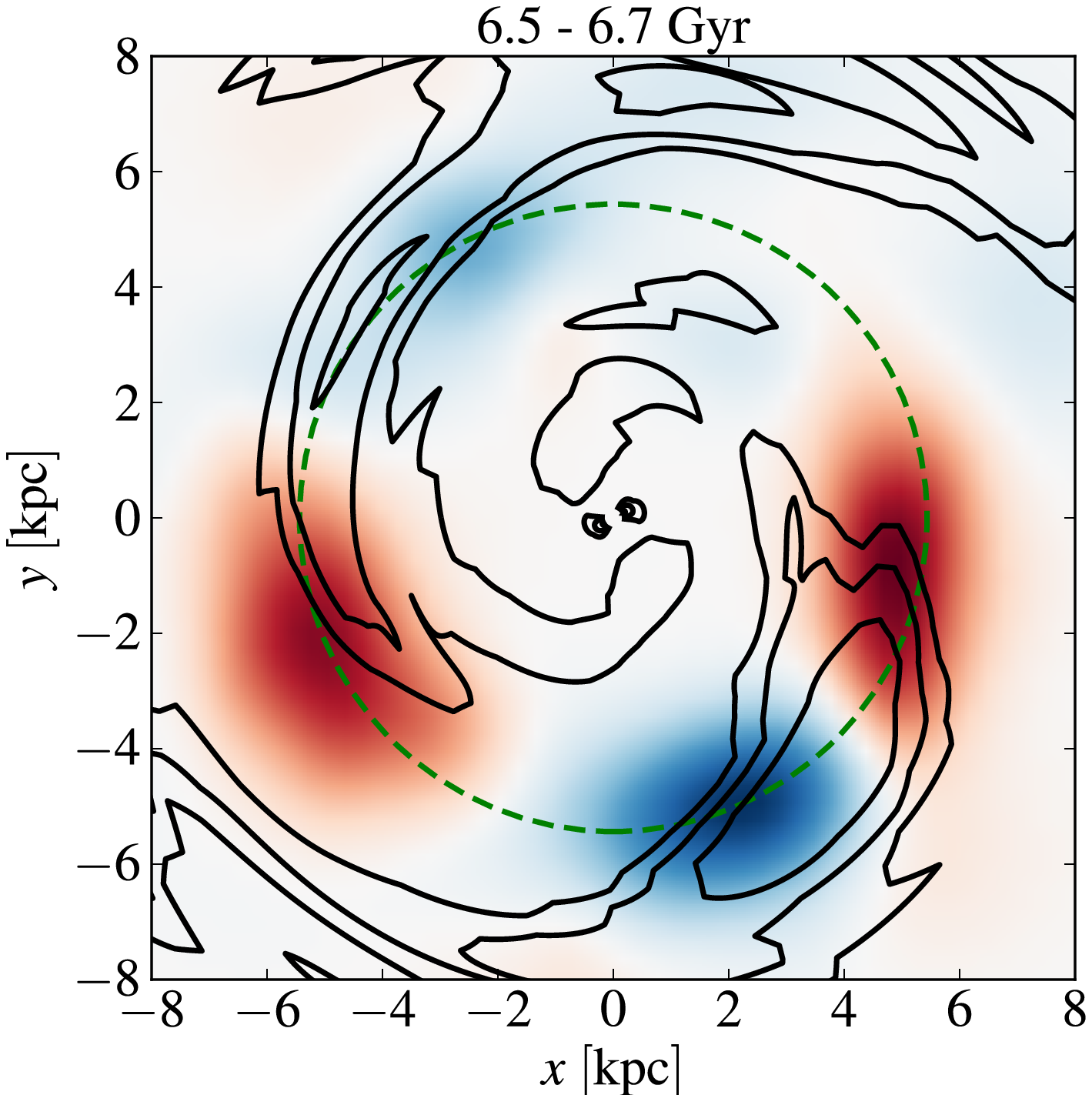}

\caption{Density of particles migrating significantly in the time
  intervals 5.3-5.5 Gye (top) and 6.5-6.7 Gyr (bottom). The sole
  selection criterion is that the particles are in the top 5 percent
  in terms of their $|\Delta j_z|$ in the given time interval. The
  blue and the red colors show the outward and inward migrators
  respectively. The black contours show the surface density of stars
  reconstructed from the $m=2$ through $m=4$ Fourier components. The
  dashed green circles mark the CR of the dominant pattern in the
  given time interval (65 km/s/kpc in the top panel and 45 km/s/kpc in
  the bottom). The direction of rotation is counter-clockwise.}
\label{fig:migrator_density}
\end{figure}

In Figure~\ref{fig:migrator_density} we demonstrate that this is
indeed the case by plotting the density of inward and outward
migrators (in red and blue respectively - darker colors indicate
higher density) for the 5.3-5.5 Gyr time interval (top panel) and
6.5-6.7 Gyr time interval (bottom panel). Spiral overdensity,
reconstructed from $m=2$ through $m=4$ Fourier components, is shown in
black contours. The first interval is chosen because it is near the
time when two of the patterns are peaking simultaneously, while during
the later time interval only the 43 km/s/kpc spiral peaks.  Note that
we did not select these particles specifically to isolate ones
interacting with either spiral. The only selection criterion is that
in the given time interval they are in the top 5 percent of
migrators. However, essentially all of the migration in both time
intervals is occuring around the CRs with most of the inward and
outward migrators respectively leading and trailing the dominant
spiral. Low densities of migrators extend to other regions of the disk
where other spirals play a role, but the majority of migration is
taking place around the 43 km/s/kpc spiral, as expected from
Figure~\ref{fig:dj_j_tot}. The dashed green circle corresponds to the
CR of the 43 km/s/kpc spiral and passes directly through the centers
of the highest density regions of migrators.

The top panel, showing the migrators for the time interval where two
of the faster patterns are peaking simultaneously shows very similar
behavior to the bottom panel where only one pattern is dominating. If
the overlap of resonances was causing chaotic evolution, it should
manifest itself during such a time interval (the dominant patterns are
$\sim 65$ and 30~km/s/kpc). Around a radius of 5~kpc in the top panel,
the density of both inward and outward migrators is saturated,
implying that the migration is even more dominated by a single pattern
in this time interval. The physical locations of the migrators are
strongly clustered around the individual overdensity peaks. In this
interval, there is also a weak oval feature in the center, driving
some angular momentum exchange in the inner part the disk (see the top
panel of Figure~\ref{fig:dj_j_tot}). The dashed green circle in this
panel corresponds to the CR of the 65 km/s/kpc spiral, which is
clearly where most of the angular momentum exchange is occuring. Thus
it is evident that even when multiple strong patterns are present near
their amplitude peaks, the majority of angular momentum exchange
occurs at very predictable locations, i.e. on either side of
overdensity peaks and at their CR. The migration in this earlier time
interval is stronger, hinting at the possibility that different
patterns may help feed each other's CR zones. However, as we show in
the following paragraphs, we are unable to find clear indications of
non-linear or chaotic evolution.

In columns 2 and 4 of Figure~\ref{fig:horseshoes} we show the
continuous wavelet transform (CWT) power scalogram of the $x$
component of the orbits shown in columns 1 and 3. We construct the
power scalogram by using the Morlet-Grossmann wavelet whose basis
``mother'' wavelet is given by
\[
\Psi_0(t) = \frac{1}{\sqrt{(2\pi)}}e^{-t^2/2\sigma^2}e^{-i\omega_0t}.
\]
See \citet{Daubechies:1990} for a thorough theoretical introduction to
wavelets and \citet{Torrence:1998} and \citet{Nener:1999} for a
practical guide. We use the approach described in the latter two for
calculating our wavelet power scalograms. The $y$-axis in the
spectrograms in Figure~\ref{fig:horseshoes} represents wavelet scale,
which is proportional to the inverse of frequency. We can therefore
follow the time evolution of dominant frequencies by inspecting the
most prominent ridges in the scalogram. Note that we plot the
logarithm of the power, so many of the smaller features are orders of
magnitude weaker than the obvious dominant ridges. Chaotic systems can
be identified with this method because the ridges in the scalogram
show significantly irregular curvature and discontinuities
\citep{Chandre:2003,Gemmeke:2008}.

\citet{Chandre:2003} show that in chaotic systems, ridges in the
scalogram become shorter, bent, and disordered. In our case, we find
in general mostly horizontal features extending across much of the
entire time interval during which migration takes place. As the stars
cross corotation and migrate in radius, the dominant frequency has to
change and this is reflected in the bend apparent in the main ridge in
all the panels. However, there is no evidence of significantly
scattered and bent ridges (see for example Figure~4 in
\citealt{Gemmeke:2008} and Figure~13 of \citealt{Chandre:2003}).

\begin{figure*}
\centering
\includegraphics[width=6in]{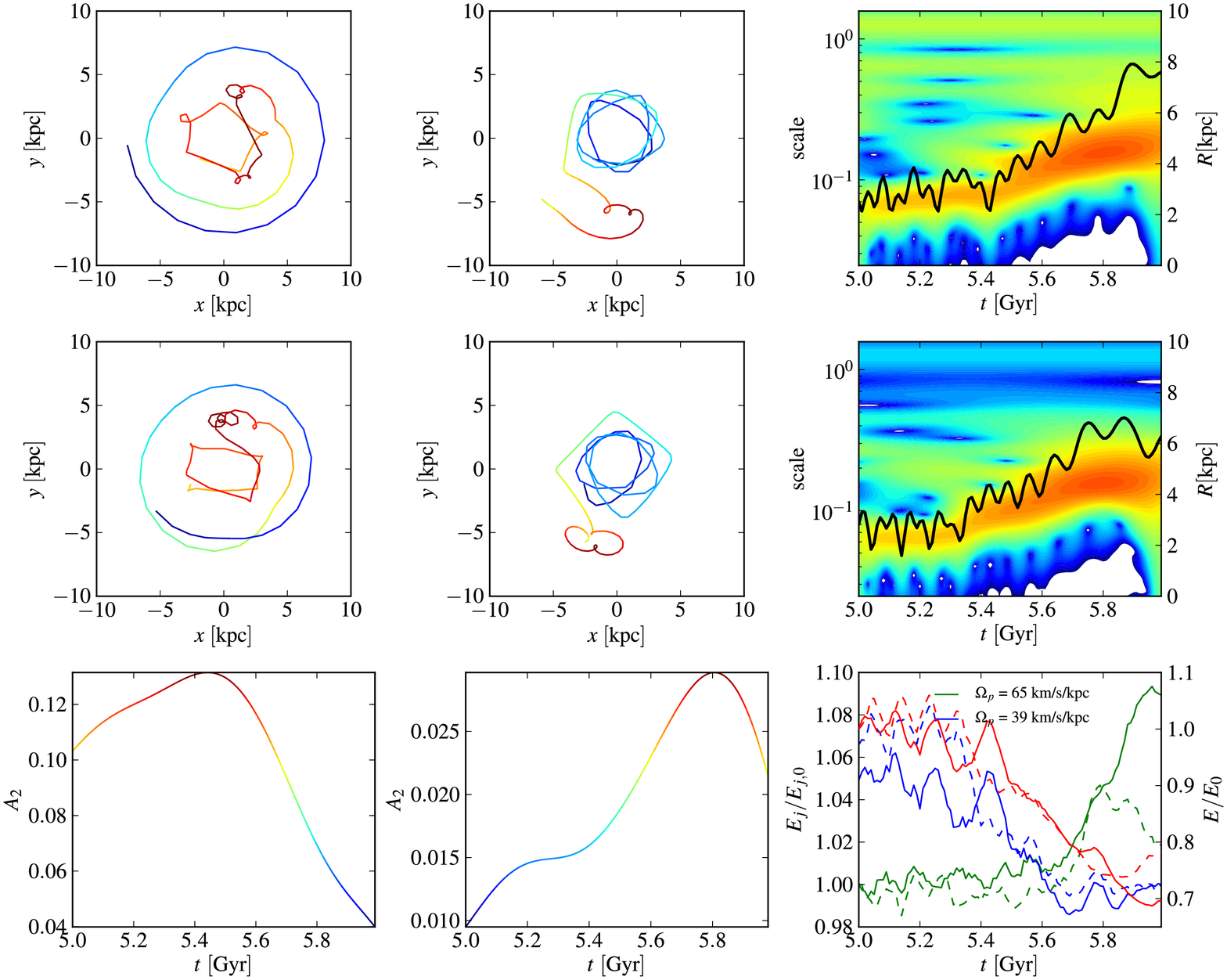}
\caption{{\bf Columns 1 and 2:} Similar to Figure~\ref{fig:horseshoes}
  but for the two orbits shown in the top row of
  Figure~\ref{fig:redist_plots}. The orbits are in the frame
  corotating with the inner spiral (column 1) and the mid-disk spiral
  (column 2) in the time interval 5-6 Gyr, when they happen to migrate
  substantially. The color corresponds to the relative spiral
  amplitude, as shown in the bottom row. All boxes measure 20 kpc in
  $x$ and $y$. The direction of rotation is counter-clockwise. Note
  that the time of CR crossing is easily identified as the orbit
  reverses direction. {\bf Column 3:} The top two rows show the
  wavelet power spectrum of the $x$-component of the orbits in color
  contours. The solid black lines show the radius $R$ in the $xy$
  plane as a function of time. The bottom panel in column 3 shows the
  fractional time evolution of the Jacobi constant, $E_j$ in the frame
  of the two spirals (green and blue lines correspond to the 65 and 39
  km/s/kpc patterns respectively) and energy $E$ (red lines). Due to
  different ranges of values, the left- and right-hand $y$-axes are
  used for $E_j$ and $E$ respectively. Particles 1 and 2 are shown in
  solid and dashed lines. When the particles are in the corotation
  region of a given pattern, $E_j$ is conserved to $\sim1\%$, while
  $E$ evolves by 30\% (though not shown, angular momentum changes by a
  factor of 2.5)}
\label{fig:horseshoes_large_dr}
\end{figure*}

The lack of obvious evidence for chaos is particularly important for
particles that migrate significantly in a short amount of time. To
investigate this further, we looked at the orbits of the two particles
in the top row of Figure~\ref{fig:redist_plots}, which migrate
$\sim5$~kpc in 0.5 Gyr. These two particles are chosen for their
extreme rates of migration, where we may also expect to detect hints
of nonlinear-effects. In the left columns of the top two rows of
Figure~\ref{fig:horseshoes_large_dr} we show the the orbits corotating
with the faster, $\sim 65$~km/s/kpc pattern, and in the middle panels
the orbits are in the frame of the $\sim 40$~km/s/kpc spiral. As in
Figure~\ref{fig:horseshoes}, the orbits are colored according to the
relative strength of the two spirals, whose amplitudes are shown in
the bottom row. The radius as a function of time for each particle is
shown as a black line in the rightmost column. Both particles cross
the CR at almost the same time near the peak of the spiral amplitude,
similar to the particles in Figure~\ref{fig:horseshoes}. The two
particles are first taken to the middle of the disk by the faster
pattern and happen to be deposited near the corotation resonance and
near the peak of the spiral pattern dominating there. They are
therefore quickly taken further outward by this second, slower
spiral. Thus, their rapid migration is simply due to two successive
corotation resonances. As before, the wavelet transform power spectrum
shown in the rightmost column of Figure~\ref{fig:horseshoes_large_dr}
shows little evidence for significantly chaotic evolution.

We look further for signs of nonlinearity by calculating the change in
the Jacobi integral for a subset of the migrators. The Jacobi
integral, $E_j = E - \Omega_p j_z$ is conserved in the rotating frame
of a single, steady perturbation \citep{Binney:2008}. In our system we
do not expect $E_j$ to be exactly conserved anywhere since multiple
patterns are always present in the disk, and the disk potential itself
is constantly changing due to accretion of fresh gas and star
formation. However, if single spirals are mostly responsible for
instantaneous changes in $j_z$ of individual stars, we would expect
that for short time intervals the distribution of $\Delta E / \Delta
j_z - \Omega_p$ to be peaked and centered at zero for particles in the
vicinity of each of the major CR regions. We select particles for
these distributions by determining the $j_{z,CR}$, i.e. the angular
momentum of circular orbits at each CR, and then selecting the top 10
percent of migrators with initial $j_z$ within 10 percent of
$j_{z,CR}$. In essence, we are selecting narrow strips of particles on
either side of each CR line shown in Figure~\ref{fig:dj_j_tot}. Note
that we determined the pattern speeds in the 6.5-6.7 case separately
from those determined in Figure~\ref{fig:frequencies} and there is a
slight discrepancy due to pattern speed evolution over Gyr timescales.

\begin{figure}
\centering
\includegraphics[width=3.2in]{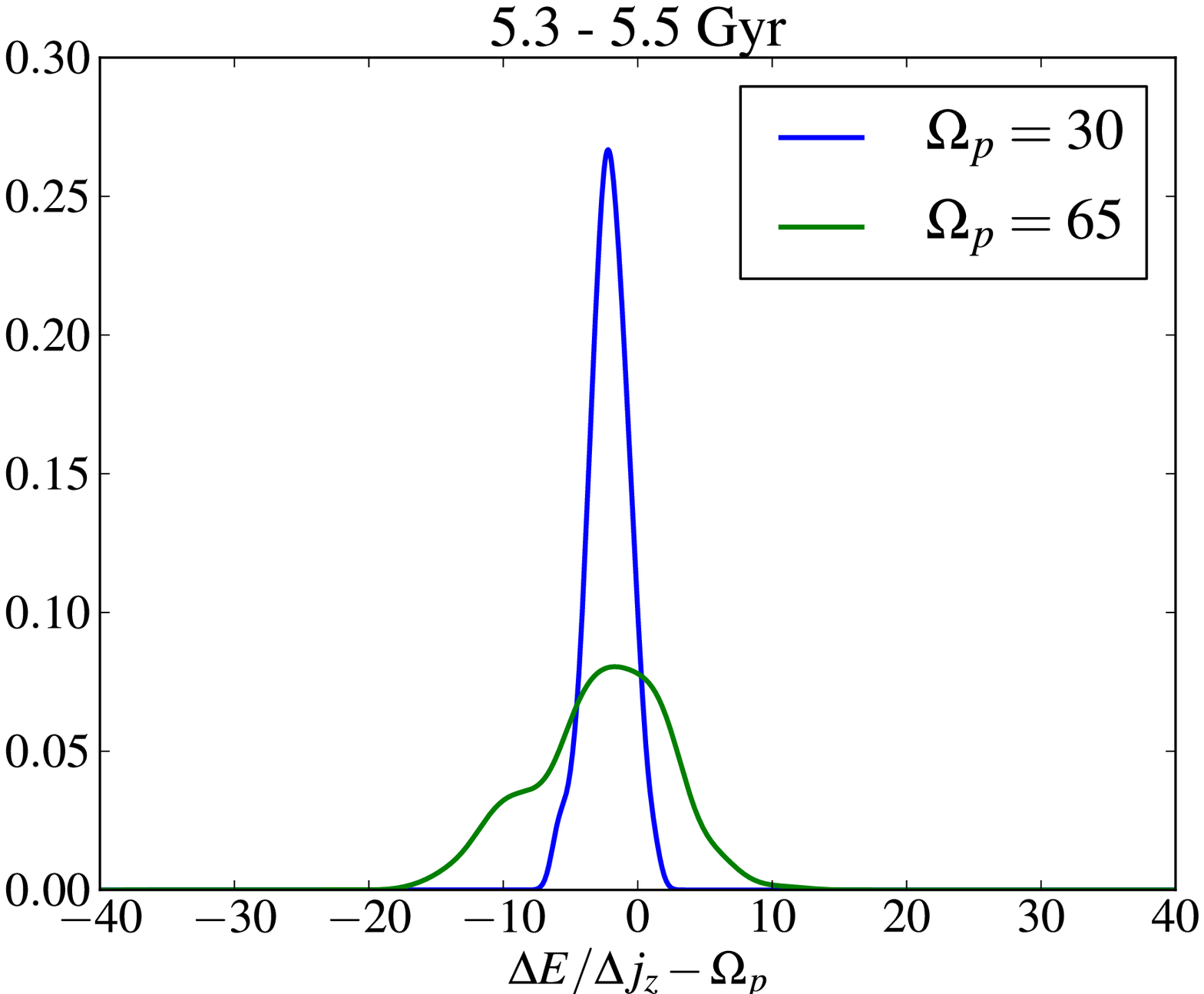}
\includegraphics[width=3.2in]{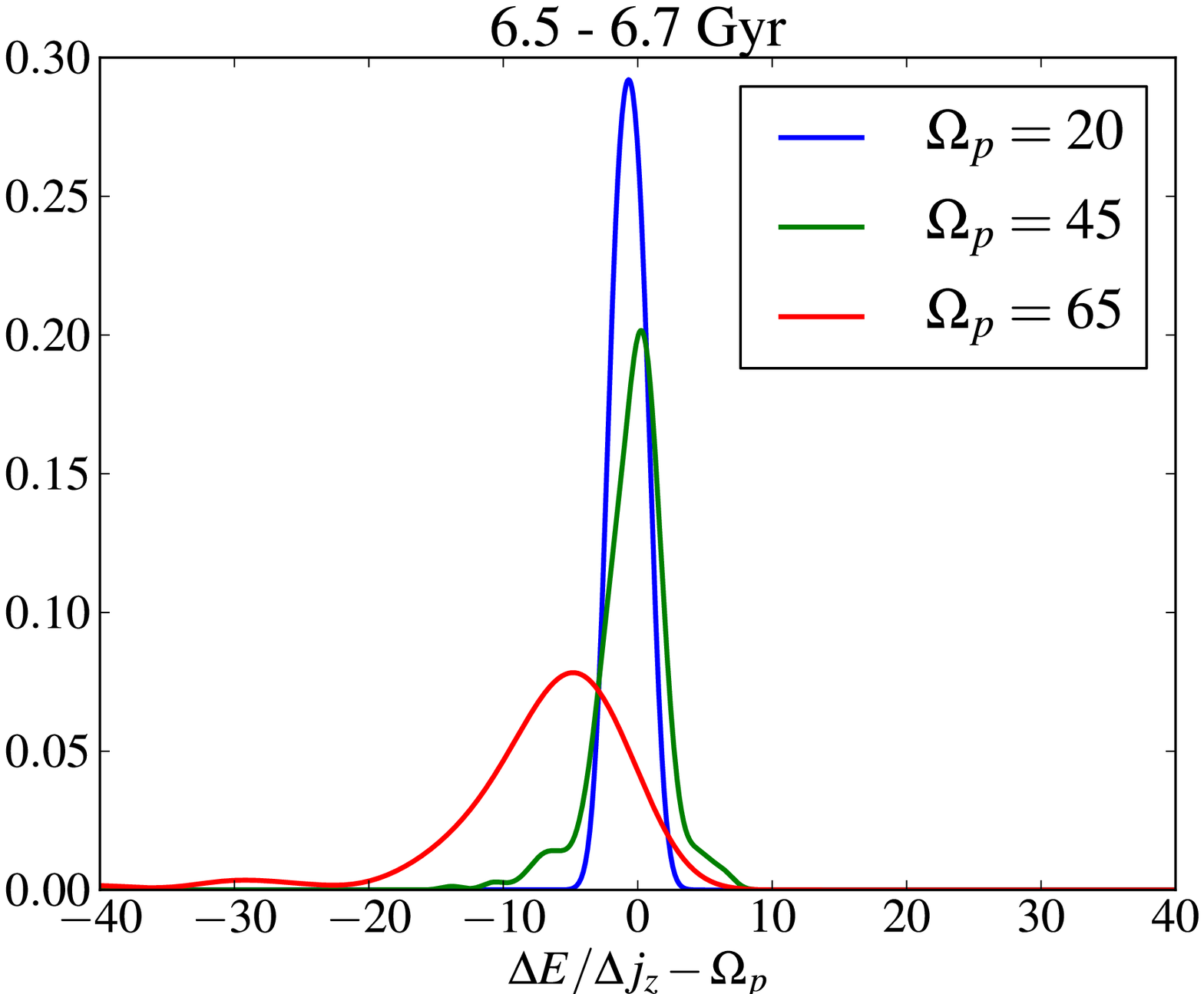}
\caption{Distributions of $\Delta E/\Delta J_z - \Omega_p$ for
  particles with the largest changes in $J_z$ near the corotation
  resonances of patterns identified from the Fourier analysis. The top
  and bottom panels show the 5.3-5.5 and 6.5-6.7 Gyr time intervals
  respectively.}
\label{fig:dedj_hist}
\end{figure}

The distributions for both time intervals are shown in
Figure~\ref{fig:dedj_hist}. All distributions are centered near zero,
within our uncertainties of pattern speed measurements of a few
km/s/kpc. The distributions for the slower perturbations dominating in
the outer disk are very strongly peaked, as expected if most of the
particles are only being affected by a single spiral. The
distributions for the faster perturbations are broadened for several
reasons. First, the spiral amplitudes are variable which broadens
their effective resonant frequencies. Second, in the inner disk other
weaker patterns may be affecting the orbits. Still, it is striking
that even in the 5.3-5.5 Gyr time interval, which was chosen
specifically because multiple strong patterns are peaking at the same
time (see also bottom panel of Figure~\ref{fig:migrator_density}), the
distributions do not appear qualitatively any different than for the
quieter 6.5-6.7 Gyr time interval. The higher density of migrators
seen in Figure~\ref{fig:migrator_density} could be due to several
factors. First, the disk density is significantly higher at the CR of
the 60 km/s/kpc spiral. The angular momentum exchange could further be
facilitated by the presence of other perturbations, essentially
feeding the CR of the dominant spiral. Nevertheless, even if the
efficiency of migration is boosted due to the presence of other
patterns, the fact that the distributions shown in
Figure~\ref{fig:dedj_hist} are centered and peaked at zero implies
that the bulk of the angular momentum exchange still proceeds due to
the CR of the individual spirals and that chaotic orbital evolution
caused by the overlap of several resonances may not be the dominant
cause of migration in this system. Note that the shift to the left
from zero for the red distribution in the bottom panel of
Figure~\ref{fig:dedj_hist} is just a few km/s/kpc and therefore well
within the measurement errors for $\Omega_p$.

Similarly, we examine the evolution of $E_j$ for the two particles
shown in Figure~\ref{fig:horseshoes_large_dr} that undergo very large
migration over a short time interval. In the bottom right panel we
show the fractional changes in the Jacobi integral for these two
particles. The fractional change for the faster pattern is measured
with respect to the beginning of the time interval, whereas for the
slower pattern we measure it with respect to the end of the interval
(this is because initially the particles interact with the faster
pattern and only later with the slower pattern). Between 5-5.6 Gyr,
$E_j$ in the frame of the 65 km/s/kpc pattern varies by $< 1\%$. In
the same time interval, the energy $E$ (shown in red) and angular
momentum (not shown for brevity) of these two particles change by 20\%
and 50\% respectively. Between 5.6-6 Gyr, $E_j$ in the frame of the
$\sim 40$~km/s/kpc pattern (blue lines) is again roughly constant,
while energy and angular momentum continue to evolve and change by
30\% and a factor of 2.5 by the end of the time interval.


%

\subsection{Orbital Circularity and Migration}

\begin{figure}

\includegraphics[width=3.2in]{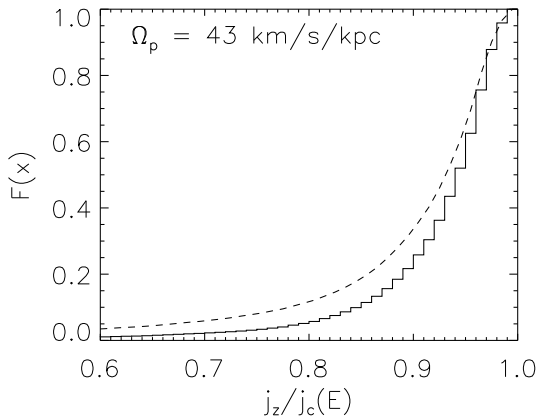}

\caption{Cumulative distribution of \jjmax weighted by $|\Delta j_z|$
  -- see equations~\ref{eq:djdist} and~\ref{eq:djcumudist} for a
  description of $F(x)$. The dashed line shows the cumulative
  distribution of \jjmax.}
\label{fig:dj_distributions}
\end{figure}

We now focus on the issue of orbital circularity and heating in CR
migration.  In Figure~\ref{fig:dj_distributions}, we show that the
amount of angular momentum exchange is directly related to the
circularity of an orbit.  This is a crucial feature of corotation
scattering, i.e. particles on the most circular orbits are also the
most susceptible to having their orbits drastically altered. It
follows that the largest changes in angular momentum are those that
are essentially kinematically untraceable, because corotation
scattering does not increase the random energy of an orbit appreciably
(\citealt{Lynden-Bell:1972}, SB02).  Particles are selected in the
same way as for Figure~\ref{fig:dedj_hist} (for brevity, we only show
the middle pattern that causes the largest amount of mixing).  From
the rotation curve and the midplane potential, we calculate the
theoretical circular orbit locus in the $(j, E)$ plane. Based on this
locus we calculate the maximum allowable angular momentum, $j_c(E)$,
of each particle based on its energy and express the circularity of
its orbit as $j_z/j_c(E)$.

We calculate the mass weighted distribution of $|\Delta j_z|$ as a
function of $x \equiv j_z/j_c(E)$ which can be expressed as
\begin{equation}
f(x) = \sum_{i=1}^N|\Delta j_{z,i}| m_i \delta(x-x_i),
\label{eq:djdist}
\end{equation}
where the subscript $i$ indicates individual particle quantities, $m$
is the mass, and $N$ is the total number of particles.  The
corresponding cumulative distribution function is given by
\begin{equation}
  F(x) = \frac{\sum_{\eta=0}^{\eta=x} f(\eta)}{\sum_{\eta=0}^{\eta=1} f(\eta)}
\label{eq:djcumudist}
\end{equation}
and shown in Figure~\ref{fig:dj_distributions}.  Note that even for
particles on fairly eccentric orbits, changes in $j_z$ are possible,
but those are most likely occurring at the Lindblad
resonances. However, the contribution of those stars to the overall
changes in $j_z$ is miniscule. The particles with \jjmax~$>0.95$
account for over 50\% of the angular momentum exchange. This follows
since $|\Delta j_z|$ is largest for those particles, but also simply
due to the fact that the disk is kinematically cool and particles on
mostly circular orbits are also most abundant. However, the mass
distribution (shown by the dashed line in the same panel) is much less
peaked than the angular momentum distribution, confirming that
particles on circular orbits are most important for the exchange of
angular momentum.

\begin{figure}
\includegraphics[width=3.5in]{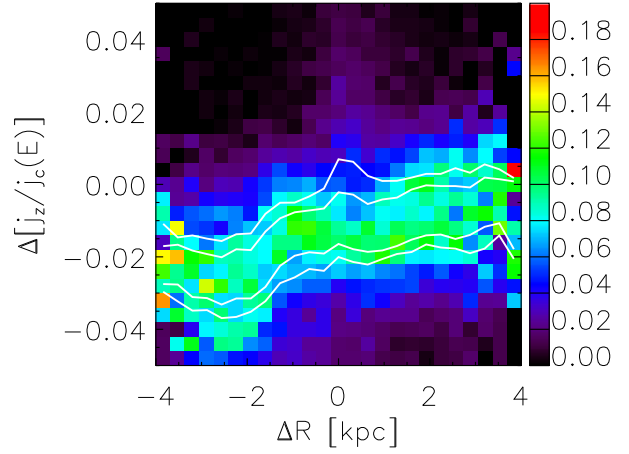}
\caption{Probability distribution of the change in the ratio of
  $j_z/j_c(E)$ for the same particles used to construct the green
  distribution in the top panel of Figure~\ref{fig:dedj_hist} and in
  Figure~\ref{fig:dj_distributions}, interacting with the
  42~\ksk~pattern over the 6.5 - 7 Gyr time interval. The inner
  (outer) white contours enclose 50\% (75\%) of the mass in each
  $\Delta R$ bin. Values are normalized such that in the vertical
  direction, $\sum P \text{d}A = 1$, where $\text{d}A$ is the area of
  the bin. }
\label{fig:dj_jc_dr}
\end{figure}

Apart from being most efficient at relocating particles on the most
circular orbits, SB02 also showed that CR scattering also does not
appreciably heat the disk. In Figure~\ref{fig:dj_jc_dr} we show a
probability density distribution of $\Delta[j_z/j_c(E)]$ given $\Delta
R$ over the same time interval studied in Figures~\ref{fig:dj_j_tot}
and~\ref{fig:dedj_hist}. Positive values of $\Delta[j_z/j_c(E)]$
indicate an \emph{increase} in circularity or a ``cooling'' of the
orbit, while negative values correspond to heating. The majority of
the particles that migrate outwards suffer very minor heating, while
the inward-migrating particles heat slightly more. The particles
moving the farthest outward also get heated the \emph{least} on
average -- this is the key feature of CR scattering because it allows
particles to experience multiple scatterings, thereby allowing for
potentially very large changes in radius during their
lifetimes. Interestingly, roughly 10\% of the particles that migrate
outward have their orbits cooled by the spiral.

Note from the top panel of Figure~\ref{fig:dj_j_tot} that the most
obvious and dramatic feature in \dj vs. $j_z$ occurs due to a
resonance with the 43~\ksk ~pattern that does not appear dominant in
Figure~\ref{fig:patspeed_sequence_fiducial}. The $\sim$20~\ksk~spiral
that dominates the power spectrum on the other hand causes relatively
little mixing. Determining the dominance of patterns from
Figure~\ref{fig:frequencies} is somewhat misleading because the power
spectrum is constructed using the normalized Fourier
coefficients. Thus, although the slowest pattern appears to dominate,
it peaks in the far outer disk where the density is low and therefore
the perturbation does not involve much mass. The middle pattern
instead is also at an optimal point in the disk. The dashed line in
Figure~\ref{fig:toomre_evol} shows the Toomre $Q$ parameter at 8 Gyr,
which is near the interval we have been analyzing. In the inner disk
where the fast pattern CR occurs, Toomre $Q$ approaches values of 4
and above - similarly in the outer disk, where the slow pattern CR
occurs Q increases rapidly and the disk density is very low (the break
occurs at $\sim8$~kpc but the CR is at $\sim11$~kpc). For the middle
pattern CR, Q is relatively low while the disk density on the other
hand is still reasonably high. Therefore the CR is well-populated by
kinematically cool stars and the middle pattern can achieve the most
mixing.

\section{Numerical Tests}
\label{sec:numerical_tests}

Spiral arms are amplified disturbances in disks, but the source of the
seed perturbation is not well understood. It is often assumed that in
isolated disks, the seed is noise in the density distribution
\citep[e.g.][]{Goldreich:1965}. In simulations like the ones presented
here, the natural source of noise is the finite particle numbers,
which are in general several orders of magnitude smaller than the
number of stars in real galactic disks, though in real galaxies giant
molecular clouds are a similar source of ``noise''. Numerical
resolution studies abound in the literature, but most often attention
has been given to the requirements of collisionless cosmological
simulations to resolve dark matter substructure
(e.g. \citealt{Moore:1998a}). In the cases where disk secular
evolution has been addressed specifically, most resolution studies
have addressed the resonant couplings between dark matter halos and
bars, but again most often only collisionless simulations were used
\citep[e.g.][]{Debattista:2000, Valenzuela:2003,
Holley-Bockelmann:2005, Weinberg:2007, Dubinski:2009a}. Spiral
structure and in particular SPH simulations in isolation are addressed
less often, though recently \citet{Christensen:2010} presented a
global disk resolution study focusing mostly on the resolution
dependence of sub-grid star formation and feedback
prescriptions. Unfortunately, their high-resolution runs use
approximately the same resolution as the fiducial runs here, because
their work was meant as a guide for cosmological simulations where the
state-of-the-art uses comparable or slightly lower resolution. Their
results therefore cannot be used to ascertain the validity of our
results.

In this section, we discuss several numerical tests to explore the
robustness and variability of the spiral activity presented in the
preceding sections. The main concern here is that the resulting
spirals may be dependent on the Poisson particle noise for their
generation and subsequent evolution. We performed runs with 0.5, 2,
and 4 times the fiducial particle number of $10^6$ particles per
component in the initial conditions. Softened gravity could be
particularly relevant to disks because it can potentially set the
relevant perturbation scales. We therefore ran several simulations
with different choices for the softening length. A possible source of
numerical heating of the disk are massive dark matter particles, so we
also ran a simulation with an order of magnitude more dark matter
particles. We discuss each of these in turn below.

\subsection{Stochasticity}

\citet{Sellwood:2009} examined the effects of stochasticity on the
growth and pattern speed evolution of bars. They found that small
perturbations in the initial conditions could lead to divergent
behavior, independent of the code used for the integration. Spirals
result from amplified disturbances and are intrinsically sensitive to
stochastic effects. It is therefore not possible to expect the pattern
speed and amplitude evolution to match exactly among the different
runs. To get some sense of the natural range of behavior due to
stochasticity, we ran two simulations with identical numerical
parameters and initial conditions, but we altered the random seed used
in the generation of initial particle positions sampled from the
distribution function. The general disk properties that result from 10
Gyr of evolution are very similar, with inner scale lengths ranging
from 3.1 to 3.3 kpc\footnote{These values were obtained using fits to
  surface density profiles. If we instead fit midplane volume-density
  profiles, the resulting scale lengths are $\sim2.5$~kpc, in
  agreement with \citet{Juric:2008} values obtained for the Milky Way
  from SDSS data.}.

\begin{figure*}
\centering

\includegraphics[width=6in]{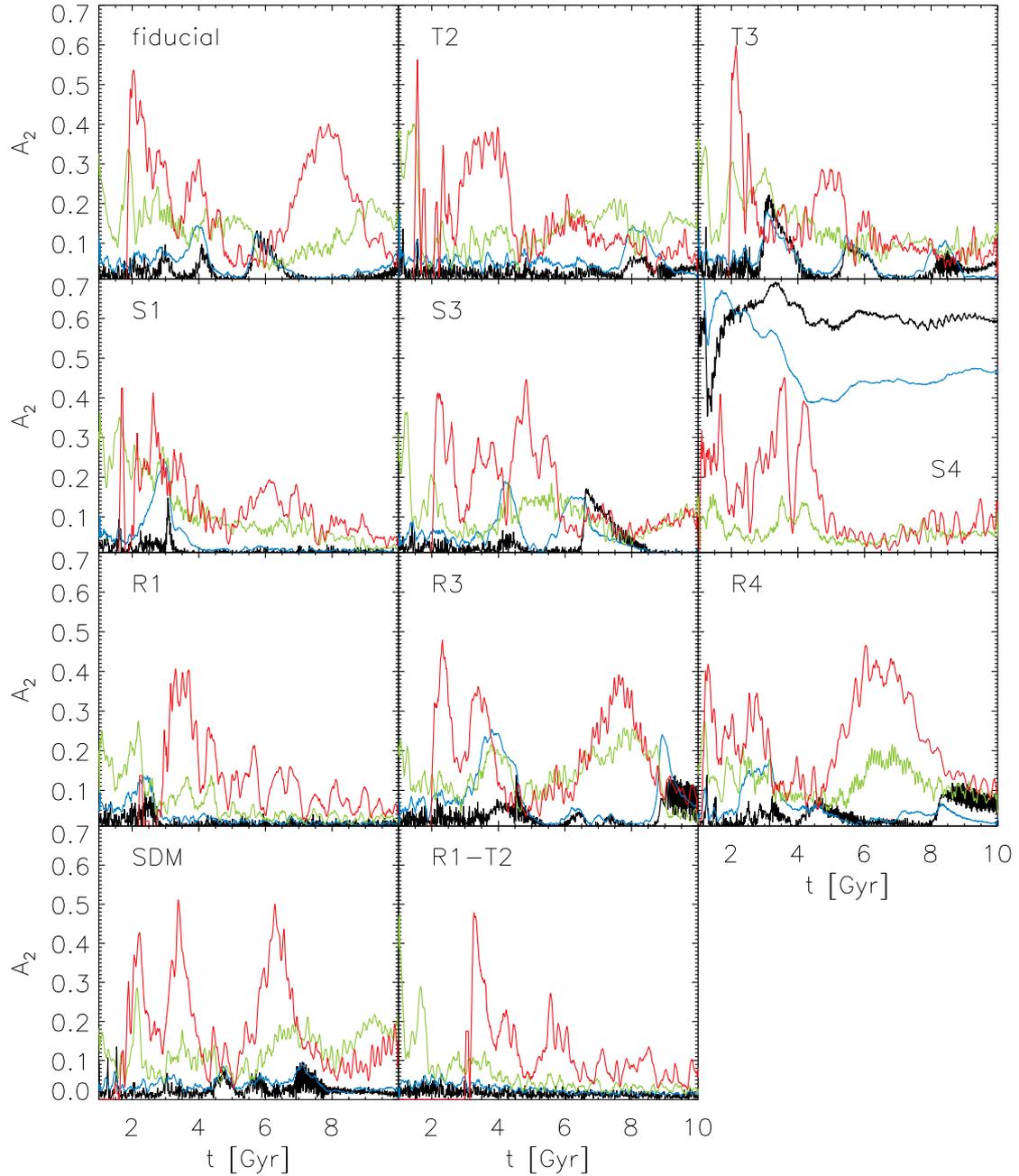}
\caption{Fourier amplitudes as a function of time for the entire
  suite. The fiducial run is shown in upper left, followed by the
  stochasticity tests in the same row. Second row shows runs with
  different softenings; the third row shows runs with different
  particle numbers. Bottom row shows runs SDM and R1-T2. Different
  color lines represent the amplitude at different radii - black,
  blue, green, and red correspond to 2, 4, 8, 12 kpc respectively.}

\label{fig:fourier_bins}
\end{figure*}

\begin{figure*}
\centering

\includegraphics[width=6in]{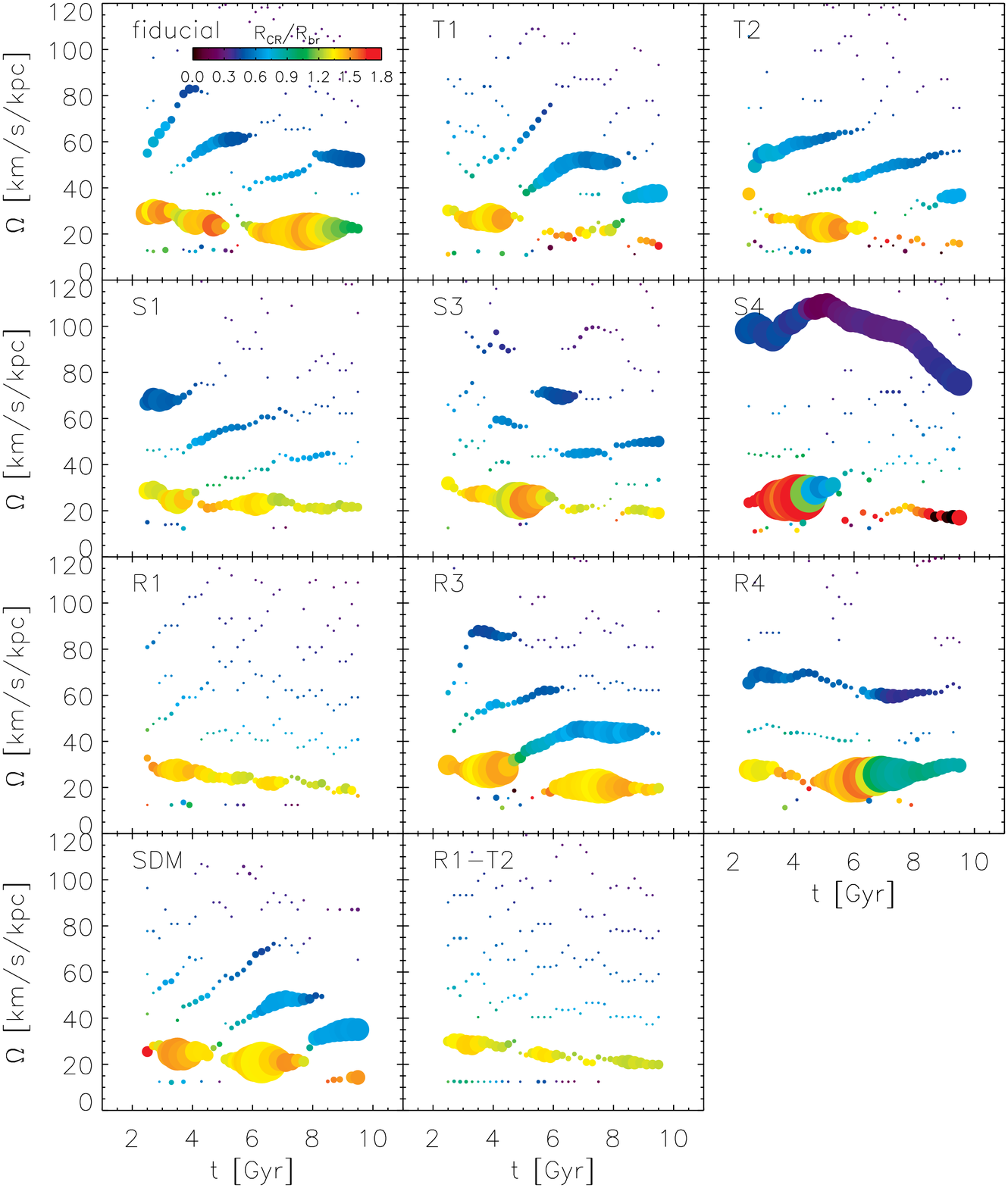}
\caption{Pattern speed evolution as a function of time. Colors and
point sizes are as in
Figure~\ref{fig:patspeed_sequence_fiducial}. Panels correspond to
panels from Figure~\ref{fig:fourier_bins}.}
\label{fig:patspeed_sequence}
\end{figure*}

In Figure~\ref{fig:fourier_bins} we show the evolution of the m=2
Fourier amplitude at four different radii in the disk for all runs. We
show smoothed m=2 time series because otherwise the rapid oscillations
of the amplitudes make it difficult to discern their overall
evolution. The colors represent the amplitude at different
radii. Bar/oval growth can be identified in this representation
whenever the black and blue lines grow together. The middle and right
panels of the top row show the stochasticity test runs (fiducial, T2
and T3) -- apparently the timing of the growth of spirals is very
different between the three runs with different random seeds. Run T2
does not grow a central oval at all (the black and blue lines showing
2 and 4 kpc are almost completely featureless), whereas the fiducial
run and run T3 both have episodes of bar/oval formation. In the
fiducial run, this occurs at $\sim3$, 4, and 6 Gyr, whereas for the T3
run it occurs at 3 and 6 Gyr. In Figure~\ref{fig:patspeed_sequence} we
show the pattern speed evolution derived in the same way as in
Figure~\ref{fig:patspeed_sequence_fiducial} with the panels
corresponding to panels in Figure~\ref{fig:fourier_bins}. Very similar
pattern speeds occur in all three disks, as shown by the top row of
Figure~\ref{fig:patspeed_sequence}. For example, the same outer (slow)
pattern can be seen in the three power spectra, starting at around
30~\ksk~and slowly decaying to 20~\ksk~by 10 Gyr. The pattern speeds
are qualitatively similar in all runs and presumably depend upon the
overall disk structure, which is relatively robust against stochastic
effects. The amplitude and timing of the spirals, on the other hand,
may vary considerably.


%


%

\subsection{Softening}

The second row of Figure~\ref{fig:fourier_bins} shows the time
evolution of the m=2 perturbations at four different radii for the
runs with different values of the softening parameter.  This suggests
that the simulations with $h_s \leqslant 100$~pc yield qualitatively
similar spiral structure, though S1 (leftmost panel) harbors somewhat
more damped spirals, especially mid-disk (green line). Run S4 is
clearly wildly different, developing a large and persistent bar very
early (in subsequent discussions we ignore this run). This comparison
shows significant variance in the timing of spiral activity, but the
amplitudes at different radii for runs S1 and S3 appear very similar
to the fiducial run (upper left panel). The second row of
Figure~\ref{fig:patspeed_sequence} shows the corresponding pattern
speed evolution and confirms that for runs S1 and S3, the pattern
speeds supported by the disk are very similar to each other and to the
fiducial run.

\subsection{Particle Number}

The third row of Figures~\ref{fig:fourier_bins}
and~\ref{fig:patspeed_sequence} show the m=2 amplitude and pattern
speed evolution for runs R1, R3, and R4 (0.5, 2, and 4 times the
fiducial particle number respectively). Run R1 stands out in this
comparison, as the structure that develops is significantly weaker. We
investigated whether this is a manifestation of stochasticity and ran
a second simulation where the initial conditions were generated using
a different random seed (run R1-T2). The results are shown in the
right panel of the bottom row - the evolution of this experiment is
essentially identical to R1, suggesting that the weaker structure is
not a stochastic effect.

The star formation rates are the same in all four runs. As a result,
the disks are comprised of proportionally similar amounts of stars --
roughly 2, 1, 4, and 8 million for the fiducial, R1, R3, and R4 runs
respectively.

In runs R3 and R4, the structure is much more similar to the fiducial
run. The timing of perturbation growth is slightly different, but the
frequencies are essentially the same for all four runs.  The
consistency of these results affirms that the runs in our simulation
suite have sufficient numbers of particles to adequately model the
disk dynamics. Ideally, we would be able to perform simulations with
still higher particle numbers to truly test for convergence, but due to
computational cost such simulations were not feasible at this time.

The ratio of CR to the break radius (i.e. the color of the points) for
the $\sim30$~\ksk~pattern in run R4 appears different from
$t\sim6.5\text{~Gyr}$ onwards. These differences are not particularly
drastic, however, and the relative locations of the CR are
consistent. The speeds of the main patterns are similar to the runs in
the top row (fiducial, T1, T2). We also see an even slower pattern,
probably due to increased particle numbers, which make its detection
in the far outer disk possible.

Finally, because dark matter particles are constantly bombarding the
disk, we explored the possibility that their perturbations may
influence the generation of spirals. In the bottom-left panel of
Figure~\ref{fig:fourier_bins}, we show the m=2 amplitude for run SDM,
which was initialized with 10 times as many DM particles. Other
properties of the run were kept the same as in the fiducial
run. Spiral amplitudes are very similar to the fiducial run, although
the innermost pattern seems to be missing. In the corresponding panel
in Figure~\ref{fig:patspeed_sequence} the frequencies as a function of
time for the run SDM appear very similar to the fiducial run. At most
times, the dominant frequencies present in both disks are essentially
the same.

While we have achieved convergence with increasing mass resolution,
the discrepancy between our lowest resolution run, R1, and the rest of
the suite is puzzling. We tested for the effect of two-body relaxation
by increasing the softening and by increasing the number of particles
in the DM halo. We also tested for potential integration issues
stemming from different softenings between the baryons and the DM
(note that in all of our runs the DM and baryons use different
softening, and this is standard practice in such simulations). Making
the baryon softening equal to the DM softening was the only experiment
that yielded results more in line with other runs.

\subsection{Overall Comparison}

\begin{figure}
\centering
\includegraphics[width=3.5in]{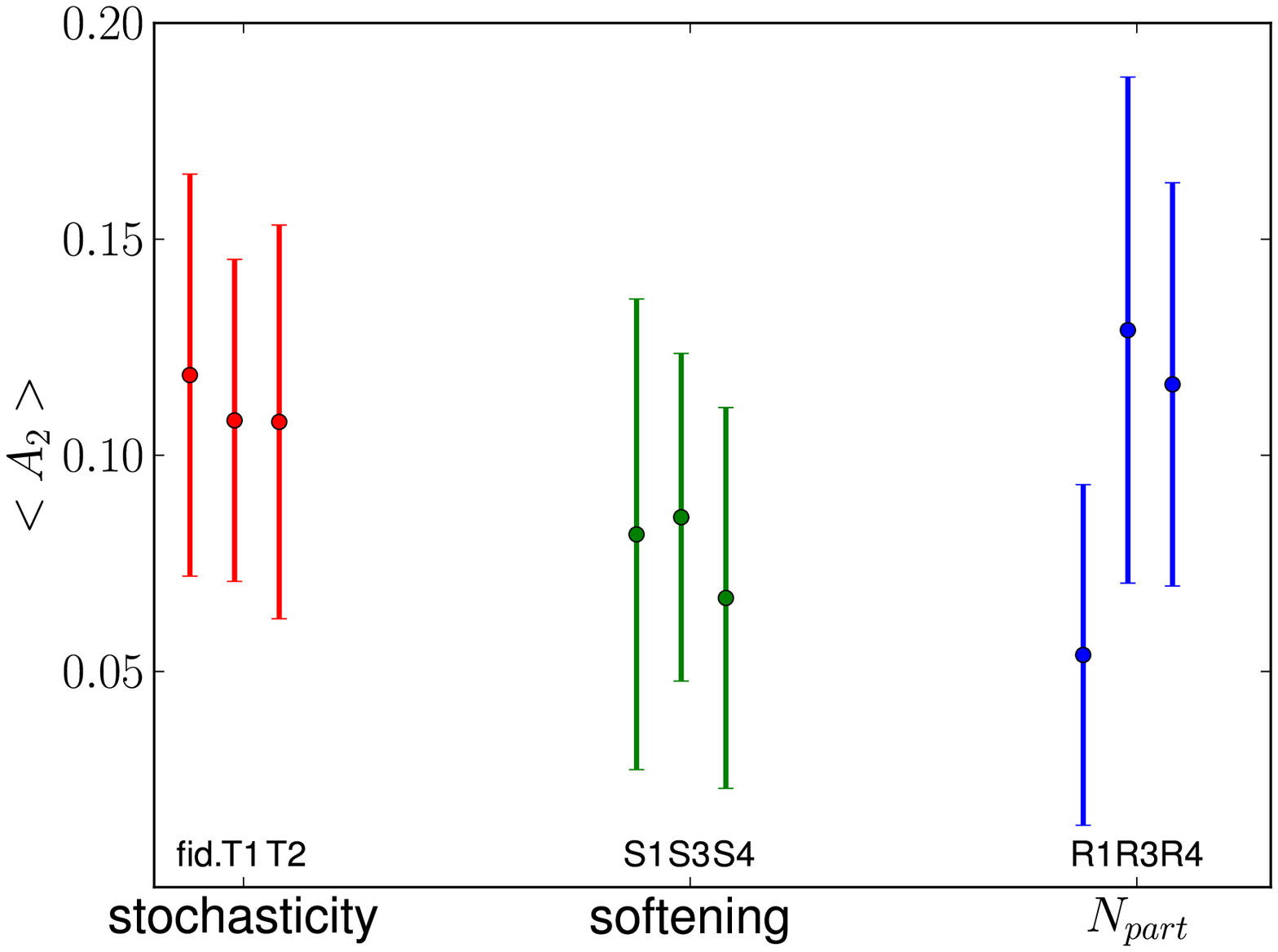}
\caption{The mean m=2 amplitude over the whole simulation for all of
  the runs in our suite. The error bars indicate the standard
  deviation. Points are offset horizontally for readability.}
\label{fig:A2_mean}
\end{figure}

In the preceding subsections we investigated the detailed evolution of
spirals in our simulation suite, but how do these simulations compare
in their global properties? In Figure~\ref{fig:A2_mean} we show the
mean m=2 amplitude, $\left<A_2\right>$, calculated over the duration
of the entire simulation, and its standard deviation represented by
the error bars. Stochasticity and choice of softening minimally impact
the global disk evolution, though all of the S-series runs have
slightly smaller $\left<A_2\right>$, especially S4. A true outlier is
R1, with lower $\left<A_2\right>$ than the rest of the suite. 

The properties of spirals are important in our understanding of radial
migration because of their effect on the distribution of stars in the
disk. We have seen in Figures~\ref{fig:fourier_bins} and
\ref{fig:patspeed_sequence} that while the range of admissible pattern
speeds is not sensitive to choices of numerical parameters, the timing
of the spirals can differ substantially. Depending on the
configuration of the disk, this could affect the migration rates and
ultimately, the predictions we extract from these models for studies
of disk stellar populations. It is difficult to quantitatively assess
the spiral structure, though we have attempted to do so in
Figure~\ref{fig:A2_mean} and found no appreciable differences in the
suite apart from run R1. However, we can quantitatively analyze the
properties of resulting stellar populations in a given region of the
disk.

\begin{figure*}
\centering
\includegraphics[width=6in]{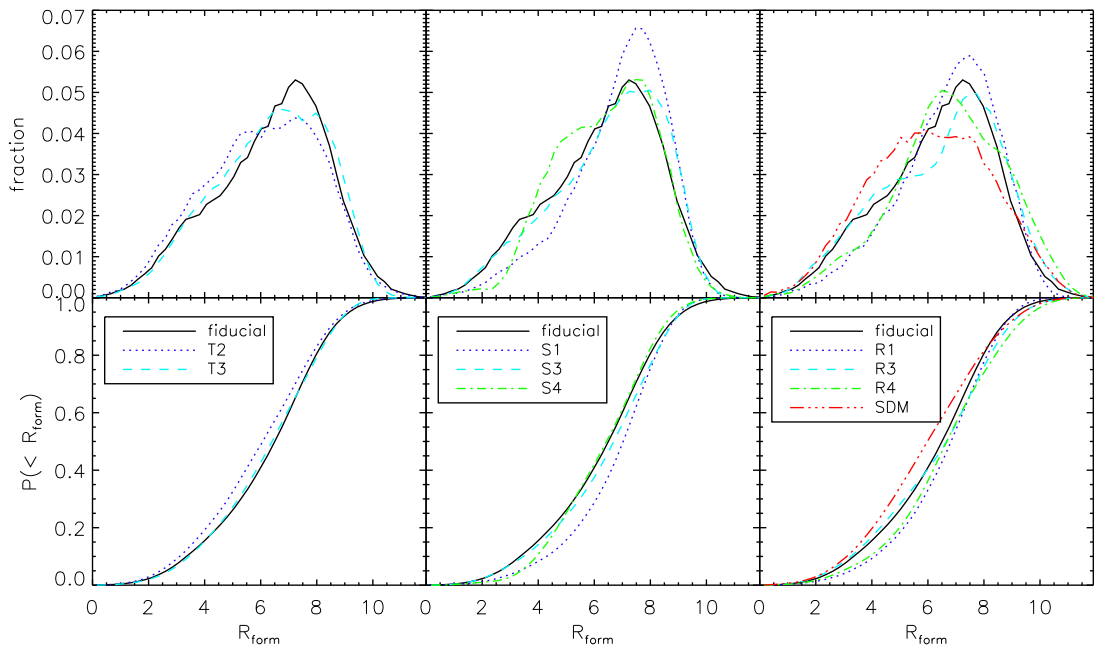}
\caption{{\bf Top:} Histograms of $R_{form}$ for particles with $7 < R_{final}$~[kpc]~$< 9$.
{\bf Bottom:} Corresponding cumulative distribution functions.}
\label{fig:rform_dist}
\end{figure*}

A natural region to examine is the model analog to the solar
neighborhood. In the top panels of Figure~\ref{fig:rform_dist} we show
the distributions of formation radii for stars that are found within
$7 < R $[kpc]$ < 9$ at the end of the simulation. This choice is
particularly important because we have addressed this same region in
our previous works (R08b, \citealt{Loebman:2011}). The bottom panels
show the corresponding cumulative distributions. The solid black line
in all panels corresponds to the fiducial run - the left, middle, and
right columns show the stochasticity, softening, and resolution tests
respectively.

The leftmost panels show that stochasticity has little overall effect
on the cumulative distributions of $R_{form}$ - at 50\% the difference
is $\leq 0.5$~kpc. Nevertheless, the distribution in the top panel
shows that the fiducial run may even over-produce the in-situ stars. 

In the remaining cumulative distributions, the overall variance at
50\% does not exceed that of the stochasticity tests. Two notable
cases are apparent - the run S1 and run SDM. Run S1, is more heavily
dominated by in-situ stars. This agrees with the fact that this run
develops weaker spiral structure in the final Gyr. However, the
appearance of the larger peak of in-situ stars for the run S1 may also
simply be another manifestation of stochasticity - when we recreate
the same distributions 2 Gyr earlier, the run S1 follows almost
exactly the fiducial distribution.

For run SDM, the peak of the $R_{form}$ distribution is actually
shifted away from the solar neighborhood. This implies even more
drastic mixing, and is indeed also apparent when we repeat the
experiment at 8 Gyr (as mentioned in the previous
paragraph). Regardless of these subtle variations, the fact that
$>50\%$ of solar neighborhood stars originated in other parts of the
disk remains robust. 

Contrary to what we might expect for run R1 given
Figures~\ref{fig:fourier_bins}, \ref{fig:patspeed_sequence} and
\ref{fig:A2_mean}, where we found that it has weaker spiral structure
on average from other runs, we find here only a small difference in
the solar neighborhood population. We would expect the resultant
migration to be much less than in the other runs. We therefore studied
the spiral structure in this run in more detail and found that while
the disk does not support strong spirals, it is permeated with weaker
and highly transient features.  We find that the transience is much
more pronounced than in the other runs. Using a 1 Gyr baseline power
spectrum, as in Figure~\ref{fig:frequencies}, we find essentially a
continuous spectrum of patterns emerges, compared with few distinct
features. This is apparent also in Figure~\ref{fig:patspeed_sequence},
where many weak patterns at a variety of pattern speeds are
identified. The pattern speeds shift on short timescales, allowing
even the weak structure to relatively efficiently redistribute the
stars.

\section{Conclusions}
\label{sec:conclusions}

Radial migration is rapidly becoming recognized as a critical process
in disk galaxies because of its wide-ranging implications. If this
process has been important for the Milky Way, it most likely played a
key role in determining the mix of stars in the solar neighborhood
today \citep{Sellwood:2002, Roskar:2008, Roskar:2008a, Schonrich:2009,
  Schonrich:2009a, Loebman:2011, Minchev:2010, Bird:2012, Lee:2011}.
On a broader scale, it has likely influenced the stellar population
gradients measured in the Milky Way disk and other galaxies
\citep{Boissier:2000, MacArthur:2004, Roskar:2008a, Williams:2009,
  Gogarten:2010, Vlajic:2011, Munoz-Mateos:2011}, and contributed
substantially to the stellar density in the outermost disk regions
(\citealt{Barker:2007, de-Jong:2007, Roskar:2008, Azzollini:2008,
  Bakos:2008, Roskar:2010, Yoachim:2010}, 2012,
in press; \citealt{Barker:2011, Alberts:2011}).

In this Paper, we have investigated in detail the origin of radial
migration by analyzing the spontaneously-forming spiral structure and
the resultant resonant angular momentum exchange. We found that the
spiral structure is transient in amplitude, but appears to support
only a few discrete pattern speeds at any given time. This means that
some stars can be tossed from the CR of one pattern to another,
resulting in large changes in radius on relatively short
timescales. Still, it is important to remember that extreme migrations
of many kpc over the course of a star's lifetime are not the norm,
they comprise the tail of the distribution. This can be seen in
Figure~\ref{fig:rform_dist} - although $\sim$50\% of the stars do come
from elsewhere, this also means that the near-majority have formed
in-situ. The situation changes with increasing radius, because the
in-situ star formation decreases - thus, the tail of the distribution
from the inner regions makes up the majority of the stellar population
at large radii. 

We demonstrated in Section~\ref{sec:angular_momentum_corotation} that
the largest angular momentum exchanges occur at the corotation of
important m=2 spirals. An important aspect of our result is the
confirmation that the largest angular momentum exchange happens for
particles on the most circular orbits - and that these particles do
not get heated by the spirals while they migrate (SB02). This is a
crucial aspect of the CR migration mechanism because it means that the
process is not self-limiting. Instead, it can continue especially for
the particles migrating the most, allowing for very large changes in
radius for some of the stars, but without betraying their journey by
anomalous kinematics. We have also searched for signs of chaotic
orbital evolution in the vicinity of resonances, but found indication
that stellar orbits remain rather regular as they migrate radially. We
have found this to be the case even for very large migrations, and
have shown that such migrations are possible in a short amount of time
if a particle passes directly from the CR of one spiral to another
(see Figures~\ref{fig:horseshoes} and~\ref{fig:horseshoes_large_dr}).

Our results show that the redistribution of stars in MW-type disks on
very short timescales is inevitable if transient spiral structure is
present. Even in the case of run R1, where the mid-disk spirals appear
to be numerically suppressed, a large fraction of stars ending up in
the solar neighborhood originated in the interior of the
disk. Therefore, our findings suggest that the effects of recurring,
spontaneous spiral structure are a key component of disk evolution
that models simply \emph{must} include if they wish to make
predictions about kinematic and chemical properties of stellar
populations. Cosmological simulations, which fail to form disks that
can support spiral structure may be missing critical aspects of disk
evolution and therefore the detailed properties of resulting disk
stellar populations must be used with care. On the other hand, the
inclusion of substructure \citep[e.g.][]{Bird:2012} is important not
only because mergers can heat and disrupt the disk, but also because
they may trigger transient structure. We hope that it should soon be
possible to use state-of-the-art cosmological simulations
(e.g. \citealt{Guedes:2011, Agertz:2011}) for detailed studies of
radial migration.

\section*{Acknowledgments}
We wish to thank the anonymous referee for constructive comments that
greatly enhanced the paper. R. R. was supported in part by the NSF ITR
grant PHY 02-05413 (also partially supporting T. R. Q.), the NASA AISR
grant NNX08AD19G, the Forschungskredit grant at the University of
Z\"urich, and the Marie Curie International Reintegration Grant.

\bibliographystyle{apj}

\end{document}